\documentclass[12pt]{article}
\usepackage{latexsym}
\usepackage{float}
\usepackage{amsmath,bm,amssymb}
\usepackage{colordvi}
\usepackage{multicol,multirow}
\usepackage{color}
\usepackage{rotating}
 \usepackage{url,hyperref}
\usepackage[ruled]{algorithm2e}
\usepackage{subfigure}
\usepackage{breqn}
\usepackage{amsfonts}
\usepackage{hyperref}
\usepackage{longtable}

\def\doublespace{\baselineskip=22pt}
\usepackage{lscape}

\usepackage[authoryear, sort&compress]{natbib}

\begin{document}
\doublespace
\baselineskip 2.8ex
\begin{center}
{\bf\Huge The Bayesian Regularized Quantile Varying Coefficient Model }

\

{\bf Fei Zhou$^{1}$, Jie Ren$^{2}$, Shuangge Ma$^{3}$ and Cen Wu$^{1}$}\\

{ $^1$ Department of Statistics, Kansas State University, Manhattan, KS}\\

{ $^2$ Department of Biostatistics and Health Data Science, Indiana University School of Medicine, Indianapolis, IN}\\

{ $^3$ Department of Biostatistics, Yale University, New Haven, CT }\\

\end{center}
%
%
{\bf $\ast$ Corresponding Author}:  Cen Wu, wucen@ksu.edu\\

\noindent {\bf\Large Abstract}\\

\noindent The quantile varying coefficient (VC) model can flexibly capture dynamical patterns of regression coefficients. In addition, due to the quantile check loss function, it is robust against outliers and heavy-tailed distributions of the response variable, and can provide a more comprehensive picture of modeling via exploring the conditional quantiles of the response variable. Although extensive studies have been conducted to examine variable selection for the high-dimensional quantile varying coefficient models, the Bayesian analysis has been rarely developed. The Bayesian regularized quantile varying coefficient model has been proposed to incorporate robustness against data heterogeneity while accommodating the non-linear interactions between the effect modifier and predictors. Selecting important varying coefficients can be achieved through Bayesian variable selection. Incorporating the multivariate spike-and-slab priors further improves performance by inducing exact sparsity. The Gibbs sampler has been derived to conduct efficient posterior inference of the sparse Bayesian quantile VC model through Markov chain Monte Carlo (MCMC). The merit of the proposed model in selection and estimation accuracy over the alternatives has been systematically investigated in simulation under specific quantile levels and multiple heavy-tailed model errors. In the case study, the proposed model leads to identification of biologically sensible markers in a non-linear gene-environment interaction study using the NHS data.   \\


\section{Introduction}

\begin{figure}[h!]
 	\centering
 	\includegraphics[angle=0,origin=c,width=1\textwidth]{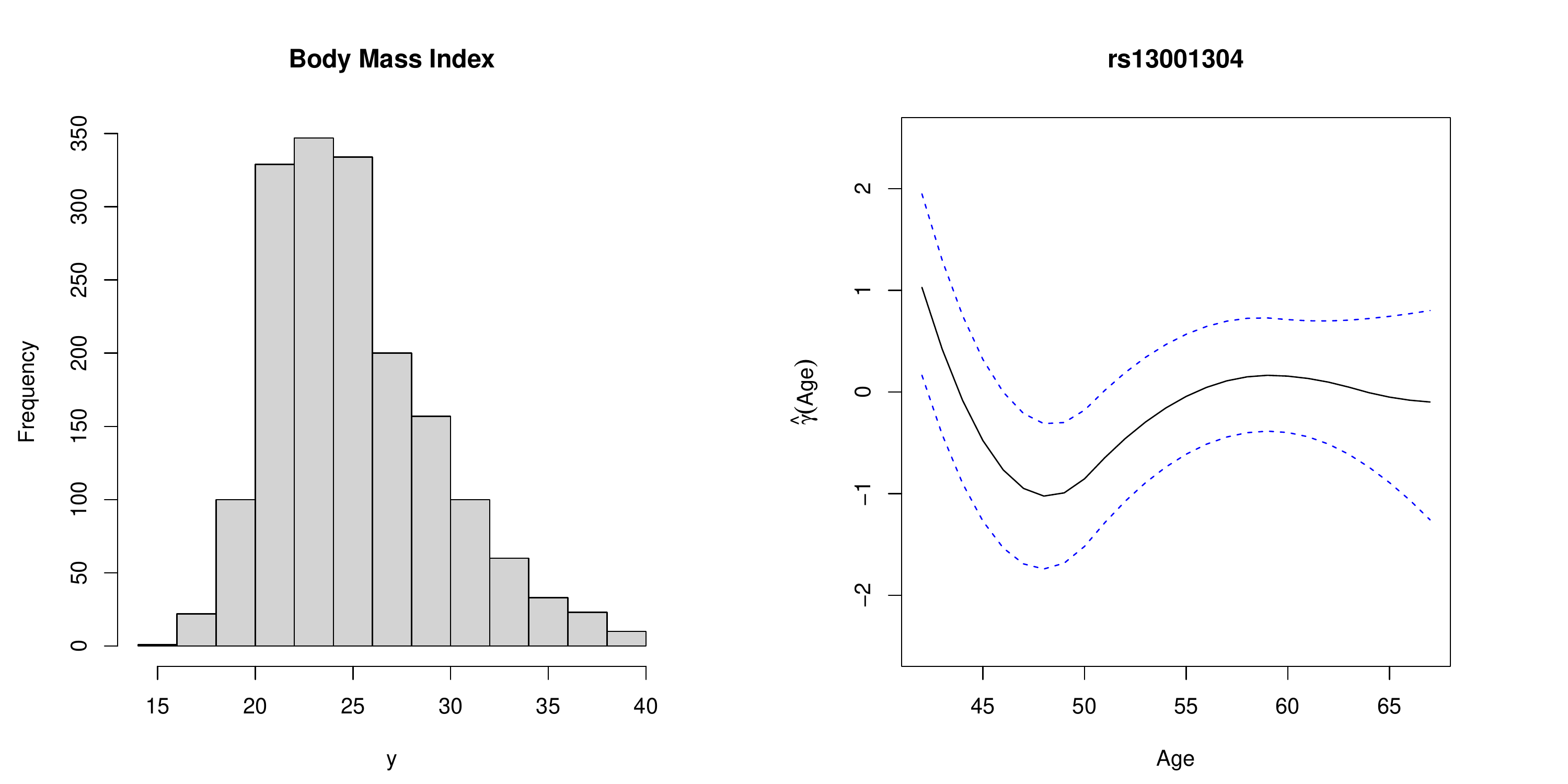}
 	\caption{Distribution of the BMI (left) and non-linear interaction effect of SNP rs13001304 (right) from the NHS data. The blue dashed lines denote the 95$\%$ credible interval.}
 	\label{fig:intro}
 \end{figure}

The quantile varying coefficient model (\cite{kim2007quantile}) has two defining characteristics. First, it can safeguard against heavy-tailed distribution and outliers due to the robustness of check loss function in quantile regression. Compared to the modeling based on conditional means, the check loss also makes a more comprehensive modeling of data feasible. Second, the quantile varying coefficient model can account for the dynamic effects of predictors on the response variable. As it has inherited from the varying coefficient model (\cite{hastie1993varying}), its regression coefficients are nonparametric functions of other variables, or effect modifiers, so the dynamic influences of the predictor can be properly captured through the varying coefficients. Therefore, the quantile varying coefficient model enjoys wide popularity and application in a broad spectrum of scientific research areas due to its robustness, superior flexibility and interpretability. For example, in the gene-environment interaction analysis (\cite{zhou2021gene}) of the Nurse's Health Data conducted in Section 5 of this paper, we aim at addressing the scientific question on how the genetic factors, which are single nucleotide polymorphisms or SNPs, are influenced by age to affect the change in body mass index (BMI). The exploratory data analysis in Figure \ref{fig:intro} clearly shows the skewness in the response variable BMI, and nonlinear interactions between SNP rs13001304 and age (the effect modifier), which justifies the use of the quantile VC model.

With a large number of the genetic factors, identification of important gene-environment interactions naturally leads to a sparse high-dimensional problem. Regularized variable selection has been extensively studied for quantile varying coefficient models. For example, \cite{Noh2012} have developed the regularization procedure based on the second order cone programming. The selection of important varying coefficients amounts to group level selection of the spline coefficients with group SCAD penalty. In longitudinal studies, \cite{Tang13} have developed adaptive LASSO based variable selection method for quantile varying coefficient models, where the group level spline coefficients are penalized via the shrinkage of the $L_v$ norm ($v\geq 1$). \cite{TangWang2012} have further examined structural identification of varying coefficients by separating the varying, nonzero constant and zero effects in quantile regression. All these studies have established the asymptotic properties of the corresponding regularized estimators in terms of (1) consistency in variable selection; that is, the proposed methods can identify nonzero quantile varying coefficient functions with probability approaching 1, and (2) the rate of convergence of the nonzero quantile varying coefficient functions. However, they have not developed the asymptotic distributions of the regularized estimators. On the other hand, \cite{dai2021inference} have established asymptotic normality and estimation consistency for a sparse kernel estimator that approximates quantile VC functions. The consistency in variable selection has not been established.

From the Bayesian perspective, variable selection for quantile varying coefficient models has not been well developed yet. One advantage of the fully Bayesian methods is that exact posterior inference can be conducted through the MCMC algorithms, even under small sample sizes. Therefore, the Bayesian analysis can provide additional insight over existing frequentist approaches, 	including the statistical inference based on credible intervals of the quantile varying coefficient functions. As the general framework for penalized (robust) variable selection can be formulated as ``(robust) loss function + penalty function" (\cite{WuMa2015,wu2019selective}), choosing the appropriate likelihood function and sparsity inducing priors, which correspond to the (robust) loss function and penalty terms respectively, have been shown to be an effective way to develop the Bayesian hierarchical models (\cite{park2008bayesian,casella2010penalized}). For Bayesian quantile regression, \cite{yu2001bayesian} have proposed using the asymmetric Laplace distribution (ALD) as the likelihood function to formulate the Bayesian quantile regression. \cite{LiXi2010} have further developed the Bayesian regularized quantile regression based on adopting the univariate and multivariate conditional Laplace priors as sparse priors. A major limitation of the conditional Laplace prior is that it does not lead to shrinkage with exact 0 coefficient, which has motivated  \cite{ren2023robust} to consider incorporating the spike-and-slab priors in bi-level selection for the Bayesian least absolute deviation (LAD) regression, a special case of the Bayesian penalized quantile regression with 50$\%$ quantile level. These methods are of a parametric nature, and cannot be adopted for analyzing the quantile varying coefficient models.

In literature, nonparametric Bayesian variable selection has been examined in varying coefficient models. \cite{LWLW} have developed Bayesian group LASSO for varying coefficient models in longitudinal studies. In gene-environment interaction studies, \cite{Ren2019} have examined the sparse structure identification for Bayesian partially linear varying coefficient models. Both work have developed Gibbs samplers for posterior sampling and inference. As the likelihood functions are employed based on normal distribution, both are not robust to long-tailed distributions and outliers in the response variable.

To the best of our knowledge, Bayesian regularized variable selection in quantile regression models with varying coefficients has not been well studied. As the quantile VC model can be further extended to a large family of non-/semi-parametric models (\cite{wang2009quantile,ma2015varying,lv2020high}), it is not feasible to investigate these models within the Bayesian framework if the cornerstone model in this family has not been fully understood from the Bayesian perspective. Therefore, to fill this gap, we have developed a novel regularized Bayesian quantile varying coefficient model. The proposed model shares the two aforementioned defining characteristics of the quantile varying coefficient model within the Bayesian framework by accommodating the heavy-tailed errors and outlying observations in the response while flexibly modeling the nonlinear interactions between the predictor and the effect modifying variable. Selection of important varying coefficients can be efficiently conducted through group level Bayesian variable selection. Incorporation of the multivariate spike and slab priors in our model promotes identification of important effects with exact sparsity, thus further improving the performance in identification and estimation. The Bayesian hierarchical model leads to a Gibbs sampler which facilitates fast posterior inference based on MCMC algorithms. We have implemented the proposed and alternative methods in R package \href{https://github.com/cenwu/pqrBayes}{pqrBayes} on the corresponding author's Github page (\url{https://github.com/cenwu/pqrBayes}). The core modules of the R package have been developed in C++. The package will be available on CRAN shortly.

\section{Statistical Methods}
\subsection{The Quantile Varying Coefficient Model}

Let $(Y_i, \boldsymbol X_i, V_i, \boldsymbol E_i), i=1,...,n,$ be independent and identically distributed random vectors, where  $Y_{i}$ is the response, $V_i$ is the univariate index variable,  $\boldsymbol X_i = (X_{i0},X_{i1},...,X_{ip})^\top$ denotes the $(1+p)$-dimensional design vector with the first element $ X_{i0}$ being 1, and $\boldsymbol E_i = (E_{i1},...,E_{iq})^\top$ is the $q$-dimensional design vector. In particular, $\boldsymbol X_i$ is of high dimensionality (e.g., denoting gene expressions), and $\boldsymbol E_i$ represents low dimensional clinical factors. At a given quantile level $0< \tau <1$, we consider the following quantile varying coefficient model: 
\begin{equation}\label{equr:vcm}
	Y_{i}=\sum_{k=1}^{q} E_{ik} \beta_{k,\tau} +\sum_{j=0}^{p}\gamma_{j,\tau}( V_i)X_{ij} +\epsilon_{i,\tau}, \quad i=1,...,n
\end{equation}
where $E_{ik}$ is the $k$th component of $\boldsymbol E_i$, $X_{ij}$ is the $j$th component of $\boldsymbol X_i$, and $\gamma_{j,\tau}(\cdot)$'s are unknown smooth varying-coefficient functions. The $\tau$th quantile of random error $\epsilon_i$ equals 0. The quantile varying coefficient model enjoys the flexibility in that the high dimensional predictors $\boldsymbol X=(\boldsymbol X_1,...,\boldsymbol X_n)^\top$ are linearly associated with the response, but the corresponding regression coefficients  $\gamma_{j,\tau}(\cdot)$'s vary with the univariate index variable $\boldsymbol V=(V_1,...,V_n)^\top$. It frequently rises in many applications that only a subset of predictors among $\boldsymbol X$ are relevant to the response variable in model (\ref{equr:vcm}), motivating the variable selection for quantile varying coefficient models. Here, $\boldsymbol E$ stands for the low dimensional clinical and environmental factors that are pre-determined as important covariates and not subject to selection. Without loss of generality, we assume that the index variable $V_i \in [0,1]$. Besides, we omit the subscript ``$\tau$" hereafter for simplicity of notation.

\subsection{The Bayesian formulation of the Quantile Varying Coefficient Model}

To formulate the Bayesian quantile varying coefficient model, we begin with approximating the varying coefficient function $\gamma_j(\cdot)$ in model (\ref{equr:vcm}) through basis expansion using polynomial splines. Denote $N_n$ as the number of uniform interior knots, and $O$ as the degree of the polynomial. Then $O$ = 1 and 2 correspond to the linear and quadratic splines respectively, and so on. Let $\boldsymbol\pi_j(\cdot)={(\pi_{j1}(\cdot),...,\pi_{jd}(\cdot))}^\top$ be a set of normalized B-spline basis with $d=N_n+O+1$ (\cite{schumaker2007spline}). Then for $j=0,...,p$, we have the following approximations
\begin{equation*}
	\boldsymbol\gamma_j(\cdot) \approx \sum_{s=1}^{d}\pi_{js}(\cdot) \alpha_{js}=\boldsymbol\alpha_j^\top \boldsymbol\pi_{j}(\cdot), 
\end{equation*}
where $\boldsymbol\alpha_j=(\alpha_{j1},...,\alpha_{jd})^\top$ is the spline coefficient vector. Subsequently, model (\ref{equr:vcm}) becomes 
\begin{equation}\label{equr:explm}
		Y_{i}=\sum_{k=1}^{q} E_{ik} \beta_k +\sum_{j=0}^{p}\boldsymbol \alpha_j ^\top\boldsymbol Z_{ij} +\epsilon_{i}.
\end{equation}
where $\boldsymbol Z_{ij}=\boldsymbol\pi_j(V_i) X_{ij}=(\pi_{j1}(V_i) X_{ij}, ..., \pi_{jd}(V_i) X_{ij})^\top$.

Given the above basis expansion, the regression coefficients $\boldsymbol \beta=(\beta_1,...,\beta_q)^\top$ and $\boldsymbol\alpha=(\boldsymbol\alpha_0^\top,...,\boldsymbol\alpha_p^\top)^\top$ can be estimated by solving the following minimization problem:

\begin{equation}\label{equr:min}
\underset{\boldsymbol\beta,\boldsymbol\alpha}{\text{argmin}}\sum_{i=1}^{n}\rho_{\tau}( Y_i-\sum_{k=1}^{q} E_{ik} \beta_k-\sum_{j=0}^{p}\boldsymbol \alpha_j ^\top\boldsymbol Z_{ij}),
\end{equation}
where $\rho_{\tau}(\epsilon_{i})=\epsilon_{i}\{\tau-I(\epsilon_{i}<0)\}$ is the check loss function for quantile regression.

Given a quantile level $\tau$, we assume that the random errors $\epsilon_i$'s from model (\ref{equr:explm}) follow an i.i.d. skewed (or asymmetric) Laplace distribution with density shown below (\cite{yu2001bayesian,yu2005three}):
\begin{equation*}
	f(\epsilon|\theta)=\tau(1-\tau)\theta \text{exp}[-\theta \rho_{\tau}(\epsilon)]=\tau(1-\tau)\theta\begin{cases}
		\scalebox{1}{$e^{-\theta\tau \epsilon}$,}& { \text{if} \; \epsilon \ge 0} \\[6pt]
		\scalebox{1}{$e^{\theta(1-\tau) \epsilon}$,}& { \text{if} \; \epsilon <0},
	\end{cases}
\end{equation*}
where $\theta^{-1}$ is a scale parameter determining the skewness of the distribution. Then the joint distribution of $\boldsymbol Y$ given $\boldsymbol E$ and $\boldsymbol Z$ can be expressed as:
\begin{equation*}
	f(\boldsymbol Y|\boldsymbol E,\boldsymbol Z,\boldsymbol\beta,\boldsymbol\alpha,\theta)=\tau^n(1-\tau)^n\theta^n \text{exp}\Big(-\theta \sum_{i=1}^{n}\rho_{\tau}(Y_i-\sum_{k=1}^{q} E_{ik} \beta_k-\sum_{j=0}^{p}\boldsymbol \alpha_j ^\top\boldsymbol Z_{ij}) \Big).
\end{equation*}
It is worth pointing out that the asymmetric Laplace likelihood is essentially a working likelihood. It has been adopted merely for the purpose to ensure that the minimization problem specified in (\ref{equr:min}) is equivalent to maximizing the above likelihood (\cite{yang2016posterior}), which allows us to work with the usual likelihood function. Because of its connection to the check loss function in quantile regression, the asymmetric Laplace distribution has been widely adopted to specify the likelihood function for Bayesian quantile regression, which sheds additional insight over the frequentist-based approaches to quantile regression.

\cite{kozumi2011gibbs} have shown that the skewed Laplace distribution can be equivalently represented as a mixture of an exponential distribution and a scaled normal distribution. To be more specific, let the random variables $u$ and $W$ be standard exponential distribution, Exp(1), and standard normal distribution, N(0,1), respectively. Define  $\kappa_1=\frac{1-2\tau}{\tau(1-\tau)}$ and $\kappa_2=\sqrt{\frac{2}{\tau(1-\tau)}}$ for $0<\tau<1$. Then we have the following representation based on a location--scale mixture of normals as
\begin{equation*}
\epsilon=\theta^{-1}\kappa_1 u+\theta^{-1}\kappa_2 \sqrt{u}W,
\end{equation*}
where $\epsilon$ follows a skewed Laplace distribution with a scale parameter $\theta^{-1}$. Consequently, model (\ref{equr:explm}) becomes

\begin{equation*}
	Y_i=\sum_{k=1}^{q} E_{ik} \beta_k +\sum_{j=0}^{p}\boldsymbol \alpha_j ^\top\boldsymbol Z_{ij}+\theta^{-1}\kappa_1u_i+\theta^{-1}\kappa_2\sqrt{u_i}W_i,
\end{equation*}
where $u_i \thicksim \text{Exp}(1)$ and $W_i\thicksim \text{N}(0,1)$. Let $\tilde{u}_i=\theta^{-1}u_i \thicksim \text{Exp}(\theta^{-1})$. Therefore, we have the following hierarchical model:
\begin{equation*}
	 Y_i=\sum_{k=1}^{q} E_{ik} \beta_k +\sum_{j=0}^{p}\boldsymbol \alpha_j ^\top\boldsymbol Z_{ij}+\kappa_1\tilde{u}_i+\theta^{-\frac{1}{2}}\kappa_2\sqrt{\tilde{u}_i}W_i.
\end{equation*}

\begin{equation*} 
	\tilde{u}_1,...,\tilde{u}_n \thicksim \prod_{i=1}^{n} \theta \text{exp}(-\theta \tilde{u}_i),
\end{equation*}
\begin{equation*}
	W_1,...,W_n \thicksim \prod_{i=1}^{n} \frac{1}{\sqrt{2 \pi}} \text{exp}(-\frac{1}{2}W_i^2).
\end{equation*}

\subsection{The Bayesian Regularized Quantile Varying Coefficient Model}

In the literature, penalized variable selection for quantile varying coefficient models have been examined with different group level penalty functions. For example,  \cite{Noh2012} have developed a group SCAD to select important groups of spline coefficients after basis expansion. \cite{Tang13} have proposed adaptive group LASSO for quantile varying coefficient models, where the group level shrinkage on spline coefficients has been imposed through the $L_{\nu}$ norm with $\nu \geq 1$. From the Bayesian perspective, the group LASSO estimator can be viewed as the posterior mode estimate when independent and identical multivariate Laplace priors are assumed for groups of regression coefficients. Such a connection has motivated us to consider the following regularized quantile varying coefficient model with group LASSO penalty: 
\begin{equation}\label{equr:glasso}
	\underset{\boldsymbol\beta,\boldsymbol\alpha}{\text{min}}\sum_{i=1}^{n}\rho_{\tau}(Y_i-\boldsymbol E_i^\top \boldsymbol\beta-\boldsymbol Z_i^\top \boldsymbol\alpha)+\lambda \sum_{j=1}^{p}||\boldsymbol\alpha_j||_2,
\end{equation}
where $||\boldsymbol\alpha_j||_2=(\boldsymbol\alpha_j^\top \boldsymbol\alpha_j)^{1/2}$, and $\lambda>0$ is the tuning parameter. We first set the independent and identical multivariate Laplace prior on $\boldsymbol\alpha_j$ as $\pi(\boldsymbol\alpha_j | \lambda,\theta) \propto (\lambda\theta)^{d}\text{exp}\{-\lambda\theta ||\boldsymbol\alpha_j||_2\}$, where $d$ is the group size (i.e. the length of $\boldsymbol\alpha_j$). The resulting posterior distribution of  $\boldsymbol\alpha$ is 
\begin{equation*}
f(\boldsymbol \alpha|\boldsymbol Y, \boldsymbol E,\boldsymbol Z,\boldsymbol\beta,\lambda, \theta) \propto \text{exp}\Big\{-\theta \sum_{i=1}^{n}\rho_{\tau}(Y_i-\sum_{k=1}^{q} E_{ik} \beta_k-\sum_{j=0}^{p}\boldsymbol \alpha_j ^\top\boldsymbol Z_{ij})-\lambda\theta\sum_{j=1}^{p}||\boldsymbol\alpha_j||_2) \Big\}.
\end{equation*}
With the reparametrization  $\eta=\lambda\theta$, the multivariate Laplace prior can be rewritten as a scale mixture of multivariate normal distribution using Gamma mixing density, that is,  

\begin{equation}\label{equr:mLap}
\setlength{\jot}{10pt}
\begin{aligned}
\text{M-Laplace}(\boldsymbol\alpha_j|\eta) &\propto {(\eta^2)}^{d/2}\text{exp}\{-\eta ||\boldsymbol\alpha_j||_2\}, \\
&\propto \int_0^{+\infty}\text{N}_d\{\boldsymbol \alpha_j|\boldsymbol 0,g_j\textbf{I}_d\}\text{Gamma}(g_j|\frac{d+1}{2},\frac{\eta^2}{2})dg_j,\\
\end{aligned}
\end{equation}

where the multivariate normal (MVN) distribution has zero mean vector and a d-by-d diagonal matrix $\text{diag}(g_j,...,g_j)$ as the covariance matrix, and the Gamma distribution is defined with the shape parameter $\frac{d+1}{2}$ and the rate parameter $\frac{\eta^2}{2}$.  By integrating out $g_j$, the conditional prior on $\boldsymbol \alpha_j$  has the multivariate Laplace distribution defined in (\ref{equr:mLap}). Therefore, the prior can be expressed as a gamma mixture of normal distributions in a Bayesian hierarchical model:
\begin{equation}\label{equr:gnmix}
\setlength{\jot}{10pt}
\begin{aligned}
\boldsymbol\alpha_j|g_j &\overset{ind}{\sim}\text{N}_d(\boldsymbol 0,g_j\textbf{I}_d),\\
g_j|\eta^2	& \overset{ind}{\sim}\text{Gamma}(\frac{d+1}{2},\frac{\eta^2}{2}),\\
\end{aligned}
\end{equation}

A major limitation of the above Laplacian shrinkage based formulation of hierarchical model is that the posterior estimates for regression coefficients $\boldsymbol\alpha$ cannot be shrunk to 0 exactly. In general, a 95\% credible interval needs to be constructed to determine the sparsity, which suffers from inaccuracy as shown in many published studies. Here, we consider incorporating multivariate spike-and-slab priors to achieve direct identification of sparsity, i.e.,

\begin{equation}\label{equr:ssprior}
	\setlength{\jot}{10pt}
	\begin{aligned}
		\boldsymbol\alpha_j|g_j,\psi_j &\overset{ind}{\sim}(1-\psi_j)\text{N}_d(\boldsymbol0,g_j\textbf{I}_d)+\psi_j\delta_0(\boldsymbol\alpha_j), \\
		\psi_j|\pi_0 &\overset{ind}{\sim}\text{Bernoulli}(\pi_0),\\
		g_j|\eta^2	& \overset{ind}{\sim}\text{Gamma}(\frac{d+1}{2},\frac{\eta^2}{2}),\\
	\end{aligned}
\end{equation}
where the spike is defined as $\delta_0(\boldsymbol\alpha_j)$, a point mass at $\boldsymbol 0_{d\times 1}$, and the slab component is  $\text{N}_d(\boldsymbol 0,g_j\textbf{I}_d)$.  The parameter $\pi_0\in[0,1]$. For $j=1,...,p$, we introduce a latent binary indicator variable $\psi_j$ corresponding to each group to conduct the selection of spline coefficients on the group level.  When $\psi_j=1$, the spline coefficient vector $\boldsymbol\alpha_j$ has a point mass density at zero, suggesting that $\boldsymbol\alpha_j$ is estimated as a zero vector and the varying coefficient corresponding to the $j$th predictor in $\boldsymbol{X}$ is 0, i.e., the $j$th predictor is not associated with the response. Besides, if $\psi_j=0$, the slab part, or the normal distribution, is in action, and the spike-and-slab prior reduces to the hierarchical priors in (\ref{equr:gnmix}), leading to a Bayesian quantile group LASSO. Therefore, $||\boldsymbol\alpha_j||_2 \neq 0$ and the $j$th group of spline coefficients is selected in final model.  By integrating out $\psi_j$ and $g_j$ in (\ref{equr:ssprior}), we have the marginal prior on $\boldsymbol\alpha_j$ as a mixture of a multivariate Laplace distribution and a point mass at $\boldsymbol 0_{d\times 1}$:
\begin{equation}\label{equr:ssMLap}
	\boldsymbol\alpha_j|\eta^2 \thicksim (1-\pi_0)\text{M-Laplace}(\boldsymbol \alpha_j|\eta)+\pi_0\delta_0(\boldsymbol\alpha_j),
\end{equation}
which borrows strength from both the Laplacian shrinkage and spike-and-slab priors. The multivariate Laplacian in the slab component plays the role as a diffuse density to model the large effects, and $\delta_0(\cdot)$ is a point mass at zero to achieve variable selection via shrinking negligible group of spline coefficients to 0. Note that (\ref{equr:ssMLap}) reduces to (\ref{equr:mLap}) when $\pi_0=0$. We assign a conjugate beta prior as $\pi_0\thicksim\text{Beta}(e,f)$ with fixed parameters $e$ and $f$, which accounts for the uncertainty in choosing $\pi_0$.

Besides, for computational convenience, we assign conjugate Gamma priors to $\eta^2$ and $\theta$ as follows:

\begin{equation*}
	\setlength{\jot}{10pt}
	\begin{aligned}
			\eta^2	& \overset{}{\sim}\text{Gamma}(c,m),\\
		\theta  &\overset{}{\sim}\text{Gamma}(a,b),\\
	\end{aligned}
\end{equation*}
where $a,b,c$ and $m$ are constants.  The multivariate normal prior has been placed on the $q$-dimensional coefficient vector $\beta=(\beta_1,...,\beta_q)^{\top}$ as:
\begin{equation*}
	\boldsymbol\beta \overset{}{\sim} \text{N}_q(\boldsymbol 0,\boldsymbol\Sigma_{\boldsymbol\beta}),
\end{equation*}	
where $\Sigma_{\boldsymbol\beta}$ denotes the covariance matrix. Similarly, for the coefficients $\boldsymbol\alpha_0$ corresponding to the varying intercept, we assign the following prior:
\begin{equation*}
\boldsymbol\alpha_0 \overset{}{\sim} \text{N}_d(\boldsymbol 0,\boldsymbol\Sigma_{\boldsymbol\alpha_0}).
\end{equation*}

\section{The Gibbs Sampler}

The joint likelihood of the unknown parameters conditional on data will be
given as
\begin{equation*}
\setlength{\jot}{10pt}
\begin{aligned}
p(\boldsymbol\alpha,\boldsymbol\beta,\tilde{u}_i,g_j,\pi_0,\theta,\eta^2|& \boldsymbol Y)\propto\\
&\prod_{i=1}^{n}\frac{1}{\sqrt{2\pi \theta^{-1}\kappa_2^2 \tilde{u}_i}}\text{exp}\{-\frac{(Y_i-\boldsymbol E_i^\top \boldsymbol\beta-\sum_{j=0}^{p}\boldsymbol \alpha_j ^\top\boldsymbol Z_{ij} - \kappa_1 \tilde{u}_i)^2}{2\theta^{-1}\kappa_2^2 \tilde{u}_i}\}\\
&\times \prod_{j=1}^{p}\Bigg((1-\pi_0) (2\pi g_j)^{-\frac{d}{2}}\text{exp}\Big(-\frac{1}{2g_j}\boldsymbol\alpha_j^\top\boldsymbol\alpha_j \Big)\textbf{I}_{(\boldsymbol\alpha_j \neq 0)}+\pi_0 \delta_0 (\boldsymbol\alpha_j)\Bigg) \\
&\times \pi_0^{e-1}(1-\pi_0)^{f-1}\\
&\times \prod_{j=1}^{p}(\frac{\eta^2}{2})^{\frac{d+1}{2}}g_j^{\frac{d-1}{2}}\text{exp}(-\frac{\eta^2}{2}g_j)\\
&\times \prod_{i=1}^{n}\theta \text{exp}(-\theta \tilde{u}_i\Big)\\
&\times \theta^{a-1}\text{exp}(-b\theta)\\
&\times (\eta^2)^{c-1}\text{exp}(-m\eta^2)\\
&\times(2\pi)^{-\frac{q}{2}}|\boldsymbol\Sigma_{\boldsymbol\beta}|^{-\frac{1}{2}}\text{exp}(-\frac{1}{2}\boldsymbol\beta^\top \boldsymbol\Sigma_{\boldsymbol\beta}^{-1}\boldsymbol\beta)\\
&\times(2\pi)^{-\frac{d}{2}}|\boldsymbol\Sigma_{\boldsymbol\alpha_0}|^{-\frac{1}{2}}\text{exp}(-\frac{1}{2}\boldsymbol\alpha_0^\top \boldsymbol\Sigma_{\boldsymbol\alpha_0}^{-1}\boldsymbol\alpha_0).
\end{aligned}
\end{equation*}
The full conditional distributions can be derived as follows. We provide all the details in the Appendix. 

$\bullet$ The full conditional distribution of $\tilde{u_i}$ is:

\begin{equation*}
\tilde{u_i}^{-1}|\text{rest}  \thicksim\text{Inverse-Gaussian}(\sqrt{\frac{\kappa_1^2+2\kappa_2^2}{(Y_i-\boldsymbol E_i^\top \boldsymbol\beta-Z_i^\top \boldsymbol\alpha)^2}},(\frac{\theta \kappa_1^2}{\kappa_2^2}+2\theta)),
\end{equation*}
 where ``rest" denotes the data and all the other model parameters sampled in the MCMC.

$\bullet$ Let $l_j=p(\boldsymbol\alpha_j=0|\text{rest})$, then the conditional posterior distribution of $\boldsymbol\alpha_j(j=1,...,p)$ is a multivariate spike-and-slab distribution given as:
\begin{equation*}
\boldsymbol\alpha_j|\text{rest} \thicksim (1-l_j)\text{N}_d(\boldsymbol\mu_j,\boldsymbol\Sigma_j) + l_j\delta_0(\boldsymbol\alpha_j),
\end{equation*}
where 

\begin{equation*}
\boldsymbol\Sigma_j=(\theta \kappa_2^{-2}\sum_{i=1}^{n}\frac{1}{\tilde{u}_i}\boldsymbol Z_{ij}\boldsymbol Z_{ij}^\top+g_j^{-1}\textbf{I}_d)^{-1},
\end{equation*}

\begin{equation*}
\boldsymbol\mu_j=\boldsymbol\Sigma_j\theta \kappa_2^{-2}\sum_{i=1}^{n}\frac{\boldsymbol Z_{ij}}{\tilde{u}_i}(Y_i-\boldsymbol Z_{i,-j}^\top\boldsymbol \alpha_{-j} -\boldsymbol E_i^\top \boldsymbol\beta-\kappa_1 \tilde{u}_i),
\end{equation*}
and
\begin{equation*}
l_j=\frac{\pi_0}{\pi_0+(1-\pi_0)|g_j \textbf{I}_d|^{-\frac{1}{2}}|\boldsymbol\Sigma_j |^{\frac{1}{2}}\text{exp}(\frac{1}{2}\boldsymbol\mu_j^\top\boldsymbol\Sigma_j \boldsymbol\mu_j )}.
\end{equation*}
Therefore, the posterior distribution of $\boldsymbol\alpha_j$ is a mixture of a multivariate normal distribution and a point mass at 0.  At each iteration of MCMC, $\boldsymbol\alpha_j$ is drawn from $\text{N}_d(\boldsymbol\mu_j,\boldsymbol\Sigma_j)$ with probability $(1-l_j)$ and is set to 0 with probability $l_j$.

$\bullet$ The full conditional distribution of $\theta$ is

\begin{equation*}
\theta|\text{rest}\thicksim \text{Gamma}\Big(\frac{3}{2}n+a, \frac{1}{2}\sum_{i=1}^{n}\frac{(Y_i-\boldsymbol E_i^\top \boldsymbol\beta-\sum_{j=1}^{p}\boldsymbol \alpha_j ^\top\boldsymbol Z_{ij}) ^2}{\kappa_2^2\tilde{u}_i} +\sum_{i=1}^{n}\tilde{u}_i+b  \Big).
\end{equation*}

$\bullet$ The full conditional distribution of $\eta^2$ is

\begin{equation*}
\eta^2|\text{rest} \thicksim\text{Gamma}\Big(\frac{(d+1)(p+1)}{2}+c, \frac{1}{2}\sum_{j=1}^{p}g_j+m\Big).
\end{equation*}

$\bullet$ The full conditional distribution of $g_j$, $j=1,...,p,$ is 

\begin{equation*}
g_j^{-1}|\text{rest} \thicksim \begin{cases}
\scalebox{1}{Inverse-Gamma($\frac{d+1}{2}$,\, $\frac{\eta^{2}}{2}$)}& { \text{if} \; \boldsymbol\alpha_{j} = 0} \\[6pt]
\scalebox{1}{Inverse-Gaussian($\sqrt{\frac{\eta^2}{\boldsymbol\alpha_j^\top \boldsymbol\alpha_j}}$,$\eta^{2}$)}& { \text{if} \; \boldsymbol\alpha_{j} \neq 0}
\end{cases}.
\end{equation*}

$\bullet$ The full conditional distribution of $\pi_0$

\begin{equation*}
\pi_0|\text{rest} \thicksim\text{Beta}\Big(1+p-\sum_{j=1}^{p}Q_j+e, \sum_{j=1}^{p}Q_j+f\Big),
\end{equation*}
where 
\begin{equation*}
Q_j=\begin{cases}
\scalebox{1}{0}& { \text{if} \; \boldsymbol\alpha_{j} = 0} \\[6pt]
\scalebox{1}{1}& { \text{if} \; \boldsymbol\alpha_{j} \neq 0}
\end{cases}.
\end{equation*}

$\bullet$ The full conditional distribution of $\boldsymbol\beta$ is multivariate normal:
\begin{equation*}
\boldsymbol\beta|\text{rest} 
\thicksim \text{N}_q (\boldsymbol\mu_{\beta^{\star}},\boldsymbol\Sigma_{\beta^{\star}}),
\end{equation*}
with covariance
\begin{equation*}
\boldsymbol\Sigma_{\beta^{\star}}=(\sum_{i=1}^{n}\frac{\theta \boldsymbol E_i \boldsymbol E_i^\top}{\kappa_2^2 \tilde{u}_i}+\boldsymbol\Sigma_{\boldsymbol\beta}^{-1})^{-1},
\end{equation*}
and mean
\begin{equation*}
\boldsymbol\mu_{\beta^{\star}}=\boldsymbol\Sigma_{\beta^{\star}}\Big(\sum_{i=1}^{n}\frac{\theta}{\kappa_2^2 \tilde{u_i}}(Y_i-\sum_{j=0}^{p}\boldsymbol \alpha_j ^\top\boldsymbol Z_{ij}-\kappa_1\tilde{u_i})\boldsymbol E_i^\top \Big)^\top.
\end{equation*}

$\bullet$ Similarly the full conditional distribution of $\boldsymbol\alpha_0$ can be obtained as
\begin{equation*}
\setlength{\jot}{10pt}
\boldsymbol\alpha_0|\text{rest} 
\thicksim \text{N}_d(\boldsymbol\mu_0,\boldsymbol\Sigma_0),
\end{equation*}
where 
\begin{equation*}
\boldsymbol\Sigma_0 = (\sum_{i=1}^{n}\frac{\theta \boldsymbol Z_{i0} \boldsymbol Z_{i0}^\top}{\kappa_2^2 \tilde{u_i}}+\boldsymbol\Sigma_{\boldsymbol\alpha_0}^{-1})^{-1}
\end{equation*}
and 
\begin{equation*}
\boldsymbol\mu_0 = \boldsymbol\Sigma_0\Big(\sum_{i=1}^{n}\frac{\theta}{\kappa_2^2 \tilde{u_i}}(Y_i-\boldsymbol E_i^\top \boldsymbol\beta-\sum_{j=1}^{p}\boldsymbol \alpha_j ^\top\boldsymbol Z_{ij}-\kappa_1\tilde{u_i})\boldsymbol Z_{i0}^\top \Big)^\top.
\end{equation*}

\section{Simulation}

We conduct a comprehensive evaluation to assess the performance of the proposed method, Bayesian regularized quantile varying coefficient model with spike and slab priors (BQRVCSS), with three alternative Bayesian methods: BQRVC, BVCSS and BVC. The BQRVC only differs from BQRVCSS in that the spike-and-slab prior is not incorporated. BVCSS and BVC are the non-robust counterpart of BQRVCSS and BQRVC, respectively. Details of the hierarchical model formulation and derivation of the corresponding Gibbs samplers are provided in the Appendix ~\ref{Appendix:Key3}. Besides, two frequentist methods, regularized varying coefficient model with adaptive group LASSO under the quantile check loss (QRVC-adp) and least square loss (VC-adp) from \cite{Tang13} are also included.

The response variable generated according to model \ref{equr:vcm} with sample size n=200 and dimensionality of $\boldsymbol{X}$ being 100 after excluding the first column of 1's. Without loss of generality, the low dimensional clinical covariates, denoted as $\boldsymbol{E}$ in model \ref{equr:vcm}, is omitted, which can facilitate a fair comparison as such a component is not included in QRVC-adp and VC-adp (\cite{Tang13}). The total dimension of regression coefficients after basis expansion is larger than the sample size. For instance, if the number of basis function is set to 5, the actual dimension is 505, including the varying intercept. The varying coefficients are set as $\gamma_0(v)=2+2\text{sin}(2\pi v), \gamma_1(v)=2\text{exp}(2v-1), \gamma_2(v)=-6v(1-v), \gamma_3(v)=-4v^3$. The rest of the coefficients are 0. We simulate two types of predictors $\boldsymbol{X}$ separately. First, the predictors are simulated from a multivariate normal distribution with mean 0 and an AR-1 covariance matrix where marginal mean is 0 and correlation coefficient is 0.5, which represents the continuous gene expression data. Second, we generate the predictors as the categorical single nucleotide polymorphism (SNP) data by dichotomizing the aforementioned gene expression values of each predictor at the 1st and 3rd quartiles, leading to the 3-level categories (0,1,2) for genotypes (aa, Aa, AA).

 We consider five error distribution for $\epsilon_{i}$'s in model (\ref{equr:vcm}): N($\mu$, 1)(Error 1), 80\%N($\mu$,1) + 20\%Normal($\mu$, 3) (Error 2), Laplace($\mu$,b) with the scale parameter b = 1 (Error 3), LogNormal($\mu$,1) (Error 4), t(2) with mean=$\mu$ (Error 5). Errors 2--5 are heavy-tailed distributions. For each error, $\mu$ is chosen so that the $\tau$th quantile is 0.  We also consider the case of non $i.i.d.$ random errors by using the following data generating model :
 \begin{equation*}
 	Y_{i}=\sum_{k=1}^{q} E_{ik} \beta_k +\sum_{j=0}^{p}\gamma_j( V_i)X_{ij} +(1+X_{i2})\epsilon_{i},
 \end{equation*}  
 where $V_i \sim \text{Uniform}(0,1)$, the $i.i.d.$ errors $\epsilon_{i}$ in model (\ref{equr:vcm}) are replaced by $(1+X_{i2})\epsilon_{i}$, and the regression coefficients are the same as in the model under $i.i.d.$ random errors.

The proportions of correct fitting (C), over-fitting (O), and under-fitting (U) are used to evaluate identification performance. In addition, the integrated mean squared error (IMSE) is adopted to assess estimation accuracy of varying coefficients. Let $\hat{\gamma} _j(v)$ denote the posterior median estimate for $\gamma_j(v)$, and $(v_1,...,v_{200})$ be the grid of points equally space on [0,1]. Therefore   $\hat{\gamma}_j(v)$ can be evaluated on the grid points $\{v_i\}_{i=1}^{200}$. Then the IMSE of $\hat{\gamma}_j (v)$ is given as IMSE$(\hat{\gamma}_j (v))=\frac{1}{200}\sum_{t=1}^{200}\big(\hat{\gamma}_j(v_t)-\gamma_j(v_t)\big)^2$. $\gamma_j(v_t)$ reduces to 0 if $j>3$. The total integrated mean squared error (TIMSE), or the sum of all the IMSE's of estimated varying coefficients,  denote the overall estimation accuracy.

We have drawn the posterior samples from the Gibbs sampler. For Bayesian methods that are based on the spike-and-slab priors, the median probability model (MPM) is adopted to identify important predictors. Define the indicator $\phi_j$ for the $j$th predictor. At the $m$th iteration, $\phi_j^{(m)}=1$ if the $j$th predictor is included in the regression model,i.e., the $j$th varying coefficient is nonzero. Then, based on M posterior samples drawn from the MCMC after excluding burn-ins, the posterior probability of including the $j$th predictor in the final model can be calculated as
\begin{equation*}
p_j=\hat{\pi}(\phi_j=1|y)=\frac{1}{M}\sum_{m=1}^{M}\phi_j^{(m)}, j=1,...,p.
\end{equation*}
A larger posterior inclusion probability suggests a stronger evidence for the importance of the corresponding varying coefficients. The MPM model consists of predictors with posterior inclusion probability no less than 0.5. It has been recommended due to its optimal prediction performance when selecting a single model is of interest (\cite{barbieri2004optimal}). For methods without using spike–and–slab priors, we use the 95\% credible interval (95\%CI) to conduct identification. In simulation, the Gibbs sampler run 10,000 MCMC iterations in which the first 5,000 samples are burn-ins.

\begin{figure}[H]
	\centering
	\includegraphics[height=9cm]{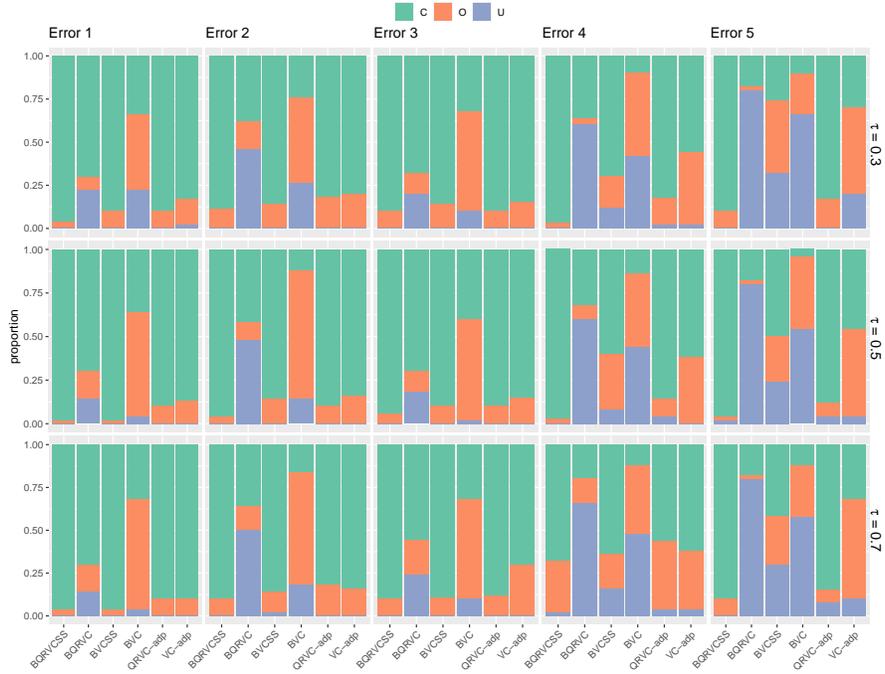}
	\caption{ Identification results for simulated gene expression data with i.i.d. errors based on 100 replicates. C: correct-fitting proportion; O: overfitting proportion; U: underfitting proportion.	\label{fig:s1}}
\end{figure}

For the 4 data generating scenarios, i.e., (1) gene expression with $i.i.d.$ error; (2) gene expression with non-$i.i.d.$ error, (3) SNPs with $i.i.d.$ error  and (4) SNPs with non-$i.i.d.$ error, all the 6 methods have been compared across 5 error distributions and 3 different quantile levels (0.3, 0.5 and 0.7). The identification results for the first scenario are shown in Figure ~\ref{fig:s1}. We can observe that under the standard normal error, BQRVCSS and BVCSS, the two Bayesian methods with the spike-and-slab priors, as well as the two frequentist methods (QRVC-adp and VC-adp), have comparable performance in correctly identifying the true model. When the random errors are heavy-tailed, Figure ~\ref{fig:s1} clearly shows the advantage of BQRVCSS over non-robust alternatives. On the other hand, BQRVCSS is apparently superior over BQRVC and BVC by yielding much larger percentage of correctly fitted models. In fact, the two Bayesian approaches without adopting spike-and-slab priors consistently lead to the two lowest proportions of identifying the true model. A comparison between BQRVCSS and QRVC-adp indicates that the two  are comparable in general, and the proposed one appears slightly better. Among the 12 sub-panels in Figure ~\ref{fig:s1}, robust methods tend to perform the worst at quantile level 0.7 under the lognormal error (Error 4), since lognormal distribution is right skewed. Such a phenomenon has not been observed under other 4 symmetric errors.

\begin{figure}[H]
	\centering
	\includegraphics[height=9cm]{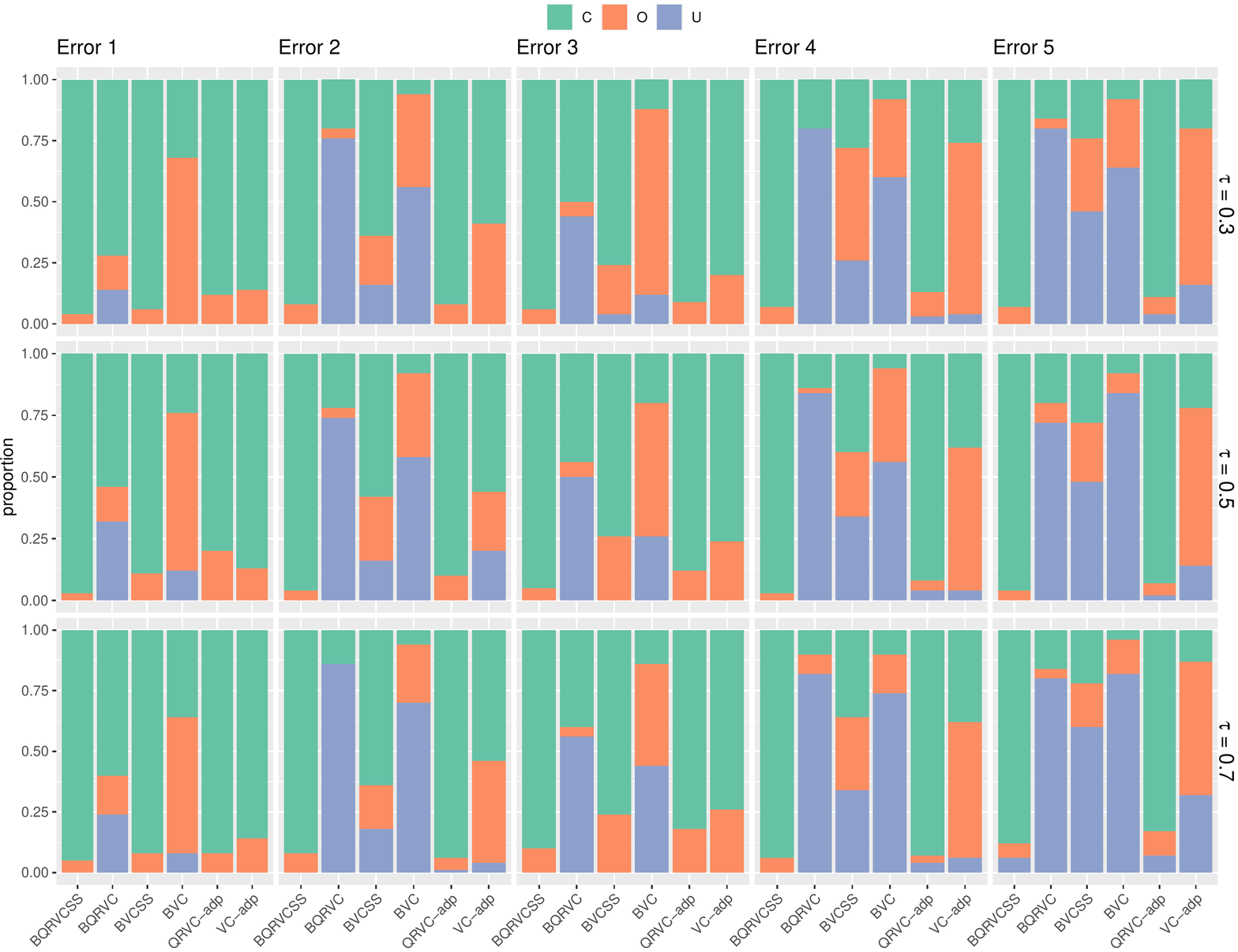}
	\caption{ Identification results for simulated gene expression data with heterogeneous errors based on 100 replicates. C: correct-fitting proportion; O: overfitting proportion; U: underfitting proportion.	\label{fig:s2}}
\end{figure}

Figure ~\ref{fig:s2} shows the identification results under the 2nd setting where the response variable is generated based on gene expression data with non-$i.i.d.$ errors. The advantage of BQRVCSS can be again concluded. Furthermore, the estimation results in terms of total integrated mean square error (TIMSE) for scenario 1 and 2 are provided in Table ~\ref{tab:s1} to Table ~\ref{tab:s2}, respectively. Under the heavy-tailed error, BVCSS leads to the smallest estimation error. For example, in Table ~\ref{tab:s1}, at quantile 0.5 with the t(2) error distribution, BQRVCSS has a TIMSE of 0.33 (sd 0.23), less than that of the BQRVC (4.35 (sd 0.78)) and QRVC-adp (0.76 (sd 0.99)), as well as non--robust alternatives. The advantage of the proposed method over the rest is due to its robustness and incorporation of the spike-and-slab prior. We also observe similar patterns in the 3rd and 4th setting from Figure ~\ref{fig:s3}, Figure ~\ref{fig:s4}, Table ~\ref{tab:s3} and Table ~\ref{tab:s4} in the Appendix.

We have also shown the estimated varying coefficients of the proposed method (BQRVCSS) for the gene expression data with $i.i.d.$ errors and 50$\%$ quantile level in the first setting in Figure ~\ref{fig:vc}. Here are the details of generating the Figure ~\ref{fig:vc}. At each replicate, a new dataset has been simulated with the aforementioned data generating model. We can obtain the posterior median estimates and 95$\%$ credible intervals after fitting the proposed method to the data generated at every replicate. The median estimates, as well as the lower and upper bound of the credible intervals, have been averaged respectively across 100 replicates to yield the estimated varying coefficients and corresponding 95$\%$ credible intervals shown in Figure ~\ref{fig:vc}. In addition, we have evaluated the empirical 95$\%$ coverage probabilities of four Bayesian methods using their pointwise 95$\%$ credible intervals over the 200 grid points. Table \ref{tab:ec} in the Appendix shows the 95$\%$ coverage probabilities for four varying coefficient functions under simulated gene expression data with i.i.d. $t$(2) errors. We can observe that overall, the proposed BQRVCSS outperforms all the alternatives. Specifically, BQRVC and BVC, the two methods not incorporating the spike-and-slab priors, can barely cover $\gamma_{1}(v)$ and $\gamma_{3}(v)$. The nonrobust counterpart BVCSS is inferior particularly at quantile level 0.3 and 0.7. The results also suggest that the performance may depend on the form of varying coefficients under estimation. It is apparent that $\gamma_{2}(v)$, a quadratic function, is corresponding to better coverage probabilities in general and none of the methods have completely missed the coverage of $\gamma_{2}(v)$, compared to those under the non-polynomial functions ($\gamma_{0}(v)$ and $\gamma_{1}(v)$) and polynomial functions with a higher order ($\gamma_{3}(v)$). \cite{yang2016posterior} have proposed a posterior variance adjustment procedure to improve the validity of credible intervals from Bayesian quantile regression with the asymmetric Laplace likelihood. While their method has been developed from a low dimensional parametric regression setting, how to adjust posterior variance to improve performance in terms of coverage probabilities in high-dimensional nonparametric setting when more complicated sparsity priors (i.e. the spike-and-slab prior) are involved worths further exploration beyond our study.

By far, the asymptotic distribution of the spline-based regularized quantile varying coefficient models have not been developed (\cite{Tang13,TangWang2012,Noh2012}). Without the asymptotic variance, it is not feasible to construct the corresponding pointwise asymptotic confidence intervals for the varying coefficients. Therefore,the counterpart of Figure ~\ref{fig:vc} for frequentist spline-based quantile VC models are not available. In literature, \cite{dai2021inference} have developed kernel-based inference procedure for estimators that approximates quantile VC in high-dimensional setting. They did not show any plots of pointwise confidence intervals for nonparametric functions. It is not immediately evident to us whether or how their methods can be used to generate confidence intervals for varying coefficient functions without the relevant specifics. Therefore we have not pursued a direct comparison to frequentist coverage of confidence intervals using their methods. For frequentist methods VC-adp and QRVC-adp, we have selected tuning parameters through Schwarz-type Information Criterion (SIC) which has been widely adopted in published literature in choosing tuning parameters for regularized (quantile) varying coefficient models (\cite{TangWang2012,Tang13,Noh2012,wang2009shrinkage}). Please refer to the Appendix for more details.

The convergence of the MCMC chains is examined by using the potential scale reduction factor (PSRF) (\cite{Gelman1992},\cite{Brooks1998}). The convergence is achieved if PSRF values are close to 1. According to \cite{gelman2013bayesian}, we use 1.1 as the cutoff (i.e. PSRF $\leq$ 1.1) to determine convergence. The PSRF has been computed for each parameter, indicating convergence of all chains after burn-ins. Figure \ref{fig:PSRF} shows the PSRF of the estimated spline coefficients of each varying coefficient function. The convergence is satisfactorily achieved.

We demonstrate the sensitivity of the proposed method BQRVCSS for variable selection to the choice of the hyperparameters for $\pi_0$ and $\eta^2$ in the Appendix and tabulate the results from Table ~\ref{tab:4-9} to Table ~\ref{tab:02}.  These results suggest that the MPM model is insensitive to different choices of the hyperparameters.  We also conduct sensitivity analysis on whether the smoothness specification of the parameters in the B spline will impact the variable selection.  The sensitivity analysis results are shown in Table \ref{tab:4-10} to \ref{tab:4-13} in the Appendix. It is evident that the proposed method is insensitive to the number of spline basis $d$, which is equivalent to $1+O+N_{n}$, in smoothness specification. We provide a heuristic justification as follows. In nonparametric literature, $n^{1/(2O+3)}$ has been established as the optimal order of number of interior spline knots under certain regularity conditions (\cite{xue2006additive}). Other orders, such as $n^{1/(2O+1)}$, has also been commonly assumed (\cite{wang2009polynomial}). Therefore, if the number of interior knots is chosen within the range of $\{\text{max}([0.5n^{1/(2O+3)}],1),[1.5n^{1/(2O+3)}]\}$, where $[a]$ denotes the integer part of $a$, the optimal order can be achieved. In practice, to avoid over fitting, cubic splines and splines with a smaller degree have been extensively used. With quadratic and cubic splines, where the spline order $O$ corresponds to 2 and 3 respectively, the aforementioned range results in 1 to 3 interior knots under the sample size 200 adopted in simulation. Therefore, the proposed method is insensitive with the above specifications of $O$ and $N_n$, i.e. the number of spline basis. Nevertheless, a rigorous justification on the optimal order of number of interior knots in high-dimensional quantile varying coefficient models remains an open question. Based on this finding, we set the degree $O=2$ and the number of interior knots $N_n=2$ for the B spline basis, which leads to $d=5$ basis functions.

The varying coefficient functions in the simulation study have been widely adopted in published nonparametric literature (\cite{Tang13,TangWang2012,Noh2012,xue2006additive,wang2009polynomial}). Functions with more complex structures may not lead to the same satisfactory performance as shown here. For example, a sine function with more oscillations in [0,1] is not a polynomial function in nature, and thus cannot be well approximated by the spline--based methods with the established optimal order of number of interior knots. We run additional simulations under setting 1 where gene expression data are generated with i.i.d. errors by only changing $\gamma_{0}(v)$ to a more complicated sine function $\gamma_{0}^{\star}(v)=2+2\text{sin}(6\pi v)$. Table \ref{tab:apendixss} in the Appendix shows that the estimation accuracy has significantly decreased for all the methods, compared to the estimation results in Table \ref{tab:s1}. In the Appendix, we have also provided the estimation plots of more complicated varying coefficient functions using the BQRVCSS and the frequentist counterpart QRVC-adp. Figure \ref{fig:vcs} and Figure \ref{fig:qrvcs} show that  $\gamma_{0}^{\star}(v)$ cannot be well modeled by both methods, which has also been observed with all the other methods (BQRVC, BVCSS, BVC and VC-adp) under comparison.

In the simulation, the figures of estimated curves are obtained based on averages over multiple replicates. To further explore the estimation performance when the proposed method has been applied to single datasets, we have also shown the figure beyond the ``average case" scenario. Specifically, at each simulation run, we compute the IMSE of posterior median estimates of the curves. Then the curves at the 25th, 50th and 75th percentile of IMSEs across all the  replicates have been overlaid with the true curve in Figure \ref{fig:vcmse} in the Appendix. We can observe that all of them are close to the true curves, although the curves at the 75th percentile of IMSEs are slightly worse than those at the other two percentiles.

\begin{table} [H]
	\def\arraystretch{1.5}
	\begin{center}
		\caption{Estimation results in terms of total
			integrated mean square error (TIMSE) for simulated gene expression data with i.i.d. errors based on 100 replicates. \\  }\label{tab:s1}
		
		\centering
		\fontsize{10}{10}\selectfont{
			\begin{tabular}{c c l c c c c c c}
				\hline
				$\tau$&&  BQRVCSS& BQRVC & BVCSS & BVC & QRVC-adp & VC-adp\\
				\hline
				$\tau=0.3$&Normal&0.23(0.10)&2.28(0.35)&0.45(0.09)&1.56(0.16)&0.25(0.10)&0.70(0.09)\\
				
				\cline{2-8}
				&NormalMix&0.34(0.19)&3.90(0.62)&0.76(0.27)&3.04(0.43)&0.45(0.23)&0.92(0.16)\\
				
				\cline{2-8}
				&Laplace&0.27(0.13)&2.97(0.45)&0.47(0.15)&2.12(0.31)&0.26(0.11)&0.71(0.11)\\
				
				\cline{2-8}
				&Lognormal&0.11(0.05)&3.38(0.55)&1.14(0.85)&5.84(1.92)&0.18(0.41)&1.22(2.45)\\
				
				\cline{2-8}
				&t(2)&0.44(0.24)&5.01(1.16)&2.63(5.24)&8.35(9.76)&0.84(0.98)&2.58(3.22)\\
				
					\hline
				$\tau=0.5$&Normal&0.21(0.06)&2.42(0.36)&0.40(0.06)&1.57(0.16)&0.21(0.07)&0.62(0.11)\\
				
				\cline{2-8}
				&NormalMix&0.31(0.17)&3.75(0.60)&0.74(0.24)&2.71(0.49)&0.35(0.16)&0.92(0.11)\\
				
				\cline{2-8}
				&Laplace&0.22(0.06)&3.07(0.48)&0.46(0.08)&1.83(0.28)&0.22(0.09)&0.70(0.08)\\
				
				\cline{2-8}
				&Lognormal&0.25(0.19)&4.59(0.94)&1.18(1.69)&5.09(2.28)&0.40(0.56)&1.26(0.68)\\
				
				\cline{2-8}
			&t(2)&0.33(0.23)&4.35(0.78)&2.04(1.48)&6.82(6.51)&0.76(0.99)&2.05(4.32)\\
			
				\hline
			$\tau=0.7$&Normal&0.21(0.08)&2.53(0.41)&0.41(0.08)&1.58(0.18)&0.23(0.10)&0.71(0.10)\\
			
			\cline{2-8}
			&NormalMix&0.33(0.14)&3.84(0.58)&0.78(0.30)&3.03(0.53)&0.45(0.26)&0.92(0.18)\\
			
			\cline{2-8}
			&Laplace&0.29(0.11)&3.22(0.49)&0.49(0.16)&2.18(0.34)&0.30(0.17)&0.73(0.12)\\
			
			\cline{2-8}
			&Lognormal&0.71(0.45)&5.44(1.52)&0.99(0.90)&4.19(2.07)&0.96(0.95)&1.35(3.65)\\
			
			\cline{2-8}
		&t(2)&0.42(0.35)&5.07(1.21)&2.65(3.35)&9.10(11.24)&0.97(1.42)&2.02(1.75)\\
		
				\hline
			\end{tabular}
		}
	\end{center}
	\centering
\end{table}

\begin{table} [H]
	\def\arraystretch{1.5}
	\begin{center}
		\caption{Estimation results in terms of total
			integrated mean square error (TIMSE) for simulated gene expression data with heterogeneous errors based on 100 replicates. \\}\label{tab:s2}
		
		\centering
		\fontsize{10}{10}\selectfont{
			\begin{tabular}{c c l c c c c c c}
				\hline
				$\tau$&&  BQRVCSS& BQRVC & BVCSS & BVC & QRVC-adp & VC-adp\\
				\hline
				$\tau=0.3$&Normal&0.35(0.15)&3.44(0.54)&0.94(0.30)&2.82(0.37)&0.37(0.20)&0.95(0.17)\\
				
				\cline{2-8}
				&NormalMix&0.50(0.24)&5.05(0.99)&1.04(1.20)&5.79(1.70)&0.45(0.23)&1.62(0.61)\\
				
				\cline{2-8}
				&Laplace&0.35(0.15)&4.04(0.79)&1.03(0.67)&3.57(0.90)&0.41(0.21)&0.94(0.27)\\
				
				\cline{2-8}
				&Lognormal&0.20(0.09)&4.18(0.93)&2.55(2.57)&9.84(4.87)&0.37(0.54)&3.59(2.03)\\
				
				\cline{2-8}
				&t(2)&0.64(0.39)&5.87(1.29)&2.99(2.83)&10.94(6.72)&1.37(1.59)&3.27(1.27)\\
				
				\hline
				$\tau=0.5$&Normal&0.27(0.21)&3.38(0.53)&0.93(0.17)&2.21(0.36)&0.28(0.16)&0.96(0.16)\\
				
				\cline{2-8}
				&NormalMix&0.29(0.12)&4.61(0.82)&1.12(0.94)&5.20(1.48)&0.35(0.16)&1.62(0.61)\\
				
				\cline{2-8}
				&Laplace&0.21(0.10)&3.84(0.67)&0.98(0.41)&3.18(0.72)&0.21(0.12)&1.06(0.33)\\
				
				\cline{2-8}
				&Lognormal&0.29(0.16)&4.36(0.95)&2.09(2.13)&8.26(3.61)&0.40(0.48)&2.45(2.17)\\
				
				\cline{2-8}
				&t(2)&0.38(0.22)&5.31(1.12)&3.33(3.15)&11.94(15.06)&1.16(2.20)&3.92(5.56)\\
				
				\hline
				$\tau=0.7$&Normal&0.33(0.11)&3.65(0.59)&0.85(0.25)&2.71(0.47)&0.38(0.16)&1.06(0.27)\\
				
				\cline{2-8}
				&NormalMix&0.51(0.22)&5.32(0.89)&1.22(1.04)&5.91(1.57)&0.78(0.56)&1.65(0.61)\\
				
				\cline{2-8}
			&Laplace&0.42(0.22)&4.25(0.73)&0.93(0.42)&3.37(0.72)&0.42(0.24)&1.10(0.39)\\
			
				\cline{2-8}
				&Lognormal&0.80(0.58)&6.85(1.71)&2.47(8.41)&7.98(6.94)&2.72(6.07)&2.54(3.39)\\
				
				\cline{2-8}
				&t(2)&0.62(0.29)&6.44(1.31)&5.37(4.67)&13.41(12.08)&1.27(1.13)&3.32(3.06)\\
				
				\hline
			\end{tabular}
		}
	\end{center}
	\centering
\end{table}

\section{Real Data Analysis}

We  analyze the Nurse's Health Study (NHS) data  from the Gene, Environment Association Studies Consortium (GENVEA) (\cite{cornelis2010gene}). The NHS aims at assessing a series of hypotheses of disease susceptibility in female based on genetic factors, i.e. single nucleotide polymorphisms (SNPs), and environmental/clinical factors in gene-environment interaction studies. The body mass index (BMI), which can quantify the obesity level, is set as the response.  We focus on SNPs on chromosome 2.  We consider age as the environment factor since it has been shown to be associated with the variations of obesity level. Besides, three clinical covariates are included: total physical activity, trans fat intake and cereal fiber intake.  The healthy subjects in the NHS are selected in the case study.  We clean the data by keeping subjects with matched phenotypes and genotypes, removing SNPs with minor allele frequency (MAF) less than 0.05 or deviation from Hardy-Weinberg equilibrium, and imputing the missing values. The final working dataset  contains 1716 subjects with 53,408 SNPs. 

A common practice in variable selection for ultra-high dimensional data in omics data analysis is to first conduct marginal screening and reduce the number of feature to a reasonable scale so (Bayesian) regularized variable selection can be applied (\cite{LWLW,WCM,wu2018additive}). Here, we screen the SNPs using the testing procedure in non-linear gene-environment interaction studies proposed by \cite{Ma2011} and \cite{WUC}.  In particular, three statistical tests have been performed to assess the effect of a genetic factor under the environmental influences and to dissect whether the interaction effects are nonlinear, linear, constant, or zero.  We keep the SNPs with p-values less than a cutoff of 0.005 from any of the tests under the response BMI.  300 SNPs pass the screening.
 
\begin{figure}[H]
	\centering
	\includegraphics[width=1\textwidth]{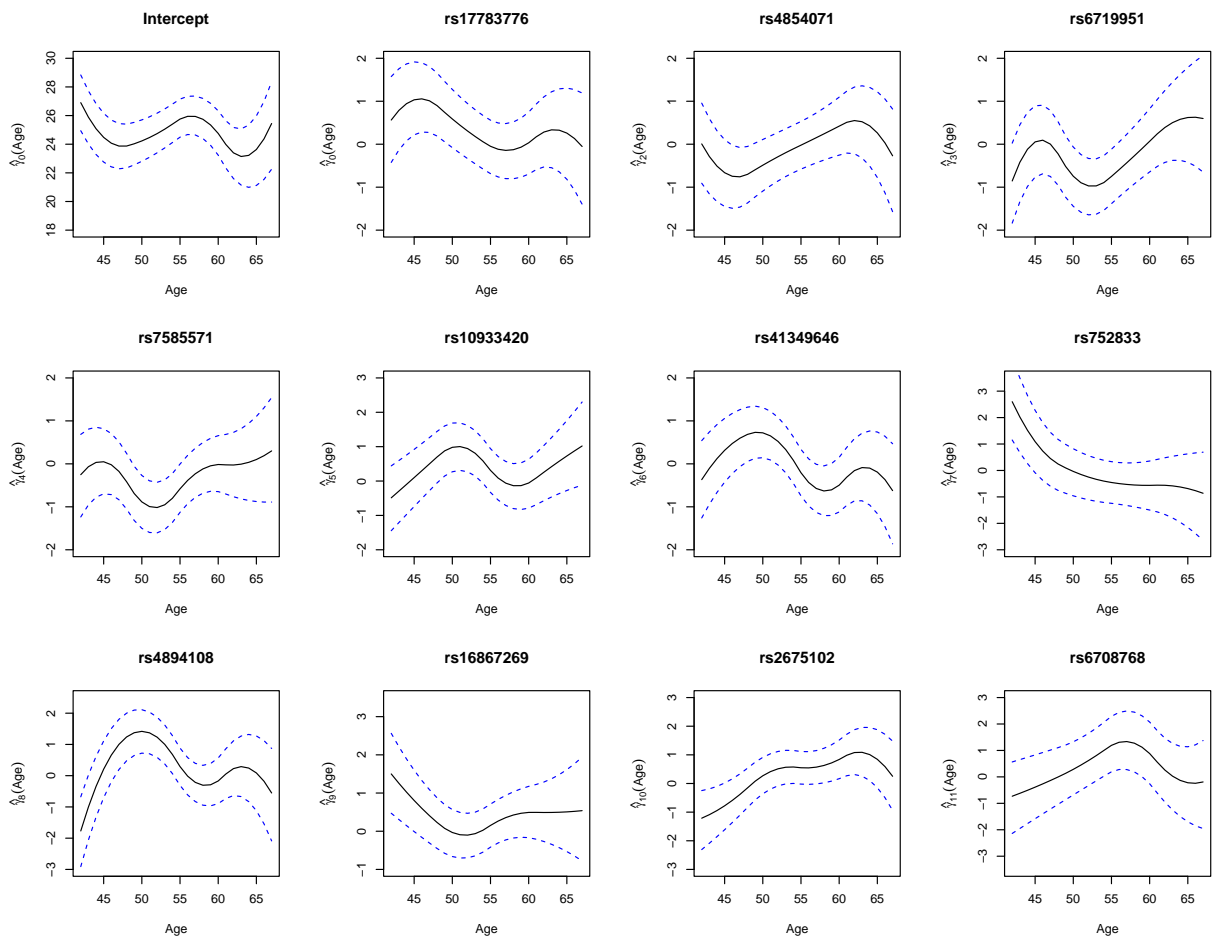}
	\caption{ Real data analysis using the proposed method (BQRVCSS). Black line: median estimates of varying coefficients for BQRVCSS. Blue
		dashed lines: 95\% credible intervals for the estimated varying coefficients.}	\label{fig:BQRVCSS}
\end{figure}

We analyze the screened data using the proposed method BQRVCSS at the median and the alternative BVCSS. Other methods, such as BQRVC and BVC are not considered since they have inferior performance in the simulation studies.  The eleven SNPs identified by BQRVCSS and the corresponding estimated varying coefficients are displayed in Figures \ref{fig:BQRVCSS}. BVCSS identifies nine SNPs which are rs17533992, rs16864365, rs6719951, rs7585571, rs752833, rs4894108, rs16867269, rs2675102 and rs13418054. Six SNPs are commonly selected by both methods.  Besides, the proposed method uniquely identified five SNPs that are located within the genes that have been reported to be associated with body weight change. For example, BQRVCSS identifies the SNP rs17783776, which is located in the gene ALK.  ALK (anaplastic lymphoma kinase) has been identified as a thinness gene which suggests it could be the target gene for obesity treatment (\cite{Orthofer2020}). As a comparison, the alternative method BVCSS misses this important gene. The proposed method also identifies rs41349646, a SNP that is mapped to the gene NPAS2.  NPAS2 has been found to play an essential role in the regulation of peripheral circadian response and hepatic metabolism, therefore affects weight change (\cite{ONeil2013}).  The SNP rs10933420 is also uniquely identified by our proposed method and it is located in the gene NGEF. \cite{Kim2015} has found NGEF associated with intra-abdominal fat accumulation. Besides, our proposed method BQRVCSS identifies rs4854071 as well.  The SNP rs4854071 is located within the gene NDUFA10 (NADH:Ubiquinone Oxidoreductase Subunit A10), which has been found to be involved in the NAFLD pathway regulating weight loss together with ten other genes (\cite{EzzatyMirhashemi2021}).

It is difficult to objectively evaluate the selection accuracy with real data. We assess the prediction performance as it may provide additional information on the performance of different methods.  We refit the selected models of BQRVCSS and BQRVC by Bayesian quantile LASSO and Bayesian LASSO, respectively, by following the refitting procedure in \cite{LWLW}.  The prediction mean squared errors (PMSEs) and prediction mean absolute deviations (PMADs) are computed based on the posterior median estimates. The proposed method BQRVCSS has the PMSE and PMAD equal to 13.13 and 1.34, respectively, while the PMSE and PMAD for BVCSS are 15.04 and 3.05, which are both larger than the counterparts of BQRVCSS. 

\section{Discussion}

Within a broader scope, regularized quantile varying coefficient model can be regarded as a robust variable selection problem in the form of ``robust loss function + penalty function" (\cite{WuMa2015}), which consists of a quantile check loss and a group level penalty function. Although other robust loss functions, including the rank based loss (\cite{wu2015penalized}), can also be considered for robust high-dimensional varying coefficient models, the regularized quantile VC model naturally leads to a Bayesian formulation if the likelihood function of the Bayesian hierarchical model is specified based on the asymmetric Laplace distribution (ALD) (\cite{yu2001bayesian}). The modeling of spline basis in the proposed study has connections to the development of semiparametric Bayesian regressions for the ``large $n$, small $p$" settings (\cite{huang2015bayesian}). As the high-dimensional Bayesian quantile VC model is underdeveloped, examining the Bayesian counterpart complements and further advances the existing studies on the quantile VC model in the frequentist framework.

Nevertheless, our limited literature search shows that high dimensional Bayesian quantile varying coefficient models have not been well examined by far. In this article, we have developed a Bayesian regularized quantile varying coefficient model. The robust asymmetric Laplace likelihood and sparsity inducing priors lead to full conditional distributions of the model parameters. Therefore, posterior inference can be efficiently conducted through Gibbs sampling. The varying coefficient model is a special case of the varying index coefficient model (VICM) when the effect modifying variable is univariate with loading weight being 1 (\cite{ma2015varying}). \cite{ma2015varying} have further shown that the new class of VICM gives rise to a broad spectrum of semi- and non-parametric models. Our study has laid a solid foundation for initiating Bayesian analyses of these models in the high-dimensional setting. Investigations on these extensions within the Bayesian framework will be postponed to the near future.

\section*{Acknowledgements}

We thank the editor, associate editor and reviewers for their careful review and insightful comments which lead to a significant improvement of this article. We also thank Yuwen Liu's help with conducting the additional simulation studies during the revision. This work was partially supported by an Innovative Research Award from the Johnson Cancer Research Center at Kansas State University and the National Institutes of Health (NIH) grant R01 CA204120.

\bibliography{BQRVC.bib}

\newpage
\clearpage
\section*{Appendix}
\appendix 

\section{Additional Simulation Results}

\subsection{Additional Identification Results}

\begin{figure}[H]
	\centering
	\includegraphics[height=9cm]{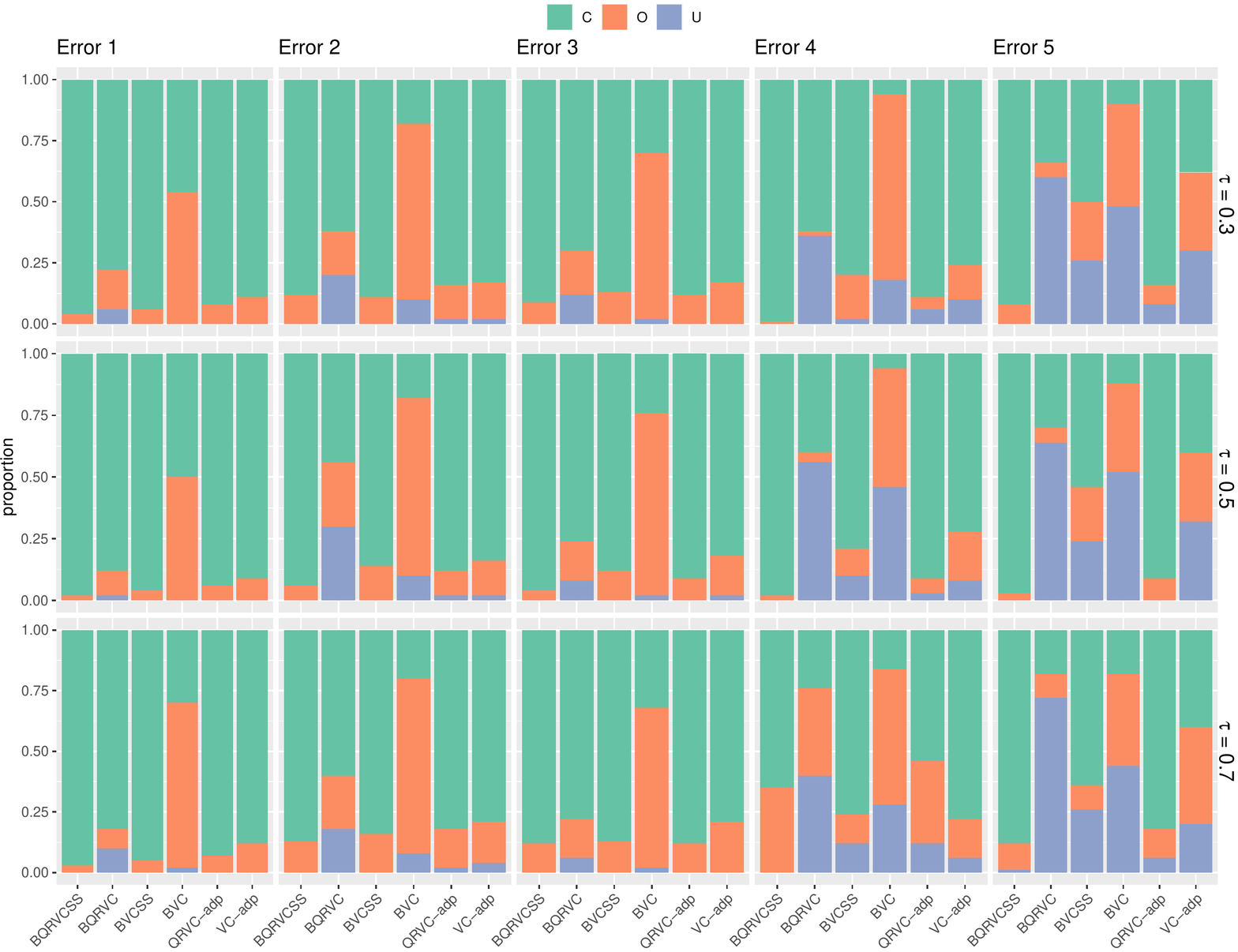}
	\caption{ Identification results for simulated SNP data with i.i.d. errors based on 100 replicates. C: correct-fitting proportion; O: overfitting proportion; U: underfitting proportion.	\label{fig:s3}}
\end{figure}

\begin{figure}[H]
	\centering
	\includegraphics[height=9cm]{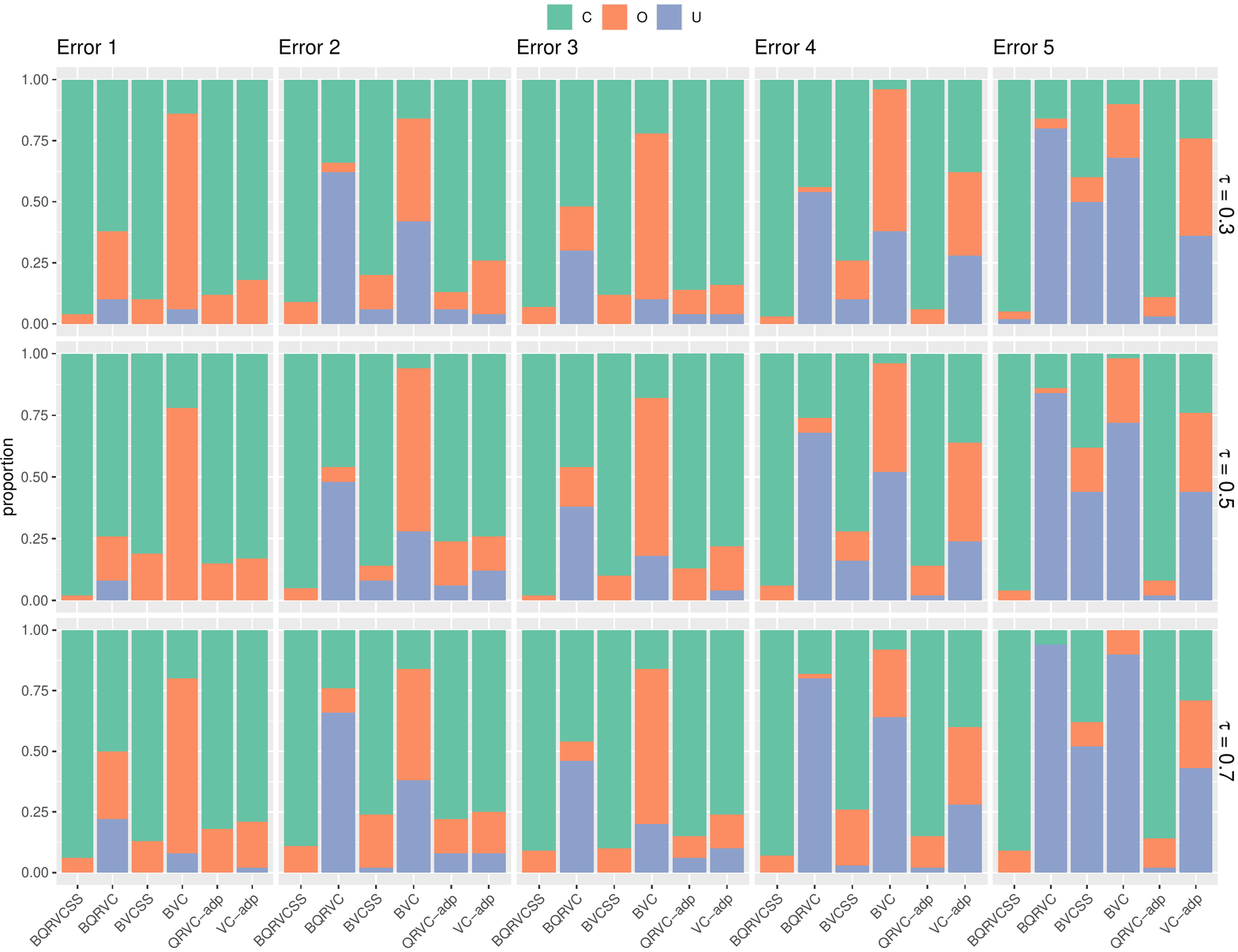}
	\caption{ Identification results for simulated SNP data with heterogeneous errors based on 100 replicates. C: correct-fitting proportion; O: overfitting proportion; U: underfitting proportion.	\label{fig:s4}}
\end{figure}

\subsection{Additional Estimation Results}

\begin{table} [H]
	\def\arraystretch{1.5}
	\begin{center}
		\caption{Estimation results in terms of total
			integrated mean square error (TIMSE) for simulated SNPs with i.i.d. errors based on 100 replicates. }\label{tab:s3}
		
		\centering
		\fontsize{10}{10}\selectfont{
			\begin{tabular}{c c l c c c c c c}
				\hline
				$\tau$&&  BQRVCSS& BQRVC & BVCSS & BVC & QRVC-adp & VC-adp\\
				\hline
				$\tau=0.3$&TMSE&0.23(0.10)&2.32(0.40)&0.45(0.12)&1.51(0.19)&0.28(0.11)&0.79(0.14)\\
				
				\cline{2-8}
				&NormalMix&0.34(0.17)&3.47(0.59)&0.76(0.23)&2.92(0.46)&0.53(0.35)&0.98(0.27)\\
				
				\cline{2-8}
				&Laplace&0.26(0.10)&2.91(0.53)&0.45(0.12)&2.06(0.32)&0.34(0.15)&0.80(0.11)\\
				
				\cline{2-8}
				&Lognormal&0.11(0.07)&3.23(0.61)&1.76(0.64)&4.7(1.38)&0.28(0.51)&1.45(0.76)\\
				
				\cline{2-8}
				&t(2)&0.38(0.17)&4.70(1.07)&1.99(1.66)&7.91(9.55)&1.30(1.30)&1.54(1.52)\\
				
				\hline
				$\tau=0.5$&Normal&0.19(0.07)&2.14(0.38)&0.41(0.09)&1.21(0.14)&0.28(0.12)&0.76(0.10)\\
				
				\cline{2-8}
				&NormalMix&0.27(0.12)&3.67(0.58)&0.73(0.16)&2.65(0.43)&0.49(0.37)&1.03(0.32)\\
				
				\cline{2-8}
				&Laplace&0.16(0.05)&2.88(0.43)&0.45(0.09)&1.87(0.35)&0.28(0.19)&0.78(0.23)\\
				
				\cline{2-8}
				&Lognormal&0.23(0.13)&4.16(0.83)&1.55(1.14)&5.3(2.43)&0.44(0.45)&1.43(0.66)\\
				
				\cline{2-8}
				&t(2)&0.31(0.18)&4.17(0.83)&1.94(1.63)&7.49(7.61)&1.25(1.23)&2.14(1.90)\\
				
				\hline
				$\tau=0.7$&Normal&0.19(0.07)&2.37(0.46)&0.41(0.10)&1.50(0.18)&0.30(0.16)&0.78(0.12)\\
				
				\cline{2-8}
				&NormalMix&0.35(0.15)&3.49(0.53)&0.7(0.19)&2.94(0.45)&0.52(0.30)&1.11(0.39)\\
				
				\cline{2-8}
				&Laplace&0.25(0.13)&2.76(0.46)&0.46(0.13)&1.99(0.27)&0.36(0.16)&0.86(0.19)\\
				
				\cline{2-8}
				&Lognormal&0.78(0.79)&5.24(1.38)&1.06(1.07)&4.21(1.91)&1.05(0.88)&0.49(0.77)\\
				
				\cline{2-8}
				&t(2)&0.46(0.38)&4.83(1.35)&1.9(1.67)&7.59(7.54)&1.13(1.01)&1.77(1.00)\\
				
				\hline
			\end{tabular}
		}
	\end{center}
	\centering
\end{table}

\begin{table} [H]
	\def\arraystretch{1.5}
	\begin{center}
		\caption{Estimation results in terms of total
			integrated mean square error (TIMSE) for simulated SNPs with heterogeneous errors based on 100 replicates.  }\label{tab:s4}
		
		\centering
		\fontsize{10}{10}\selectfont{
			\begin{tabular}{c c l c c c c c c}
				\hline
				$\tau$&&  BQRVCSS& BQRVC & BVCSS & BVC & QRVC-adp & VC-adp\\
				\hline
				$\tau=0.3$&Normal&0.26(0.11)&3.17(0.57)&0.83(0.24)&2.71(0.44)&0.35(0.24)&1.13(0.30)\\
				
				\cline{2-8}
				&NormalMix&0.40(0.20)&4.59(0.79)&1.72(0.86)&5.66(1.35)&0.63(0.54)&1.63(0.57)\\
				
				\cline{2-8}
				&Laplace&0.30(0.14)&3.72(0.68)&0.90(0.36)&3.74(0.80)&0.42(0.45)&1.17(0.49)\\
				
				\cline{2-8}
				&Lognormal&0.17(0.08)&3.54(0.67)&3.70(2.01)&8.86(3.66)&0.72(0.98)&4.32(3.00)\\
				
				\cline{2-8}
				&t(2)&0.66(0.65)&5.92(1.54)&4.64(5.71)&16.16(23.82)&2.09(4.68)&3.78(4.41)\\
				
				\hline
				$\tau=0.5$&Normal&0.17(0.08)&3.11(0.48)&0.82(0.21)&2.10(0.29)&0.25(0.18)&1.09(0.31)\\
				
				\cline{2-8}
				&NormalMix&0.25(0.12)&4.2(0.71)&1.66(0.63)&4.56(1.02)&0.68(0.73)&1.74(0.71)\\
				
				\cline{2-8}
				&Laplace&0.18(0.12)&3.78(0.63)&0.46(0.26)&3.18(0.55)&0.23(0.16)&0.85(0.45)\\
				
				\cline{2-8}
				&Lognormal&0.17(0.08)&4.20(0.74)&2.88(4.66)&9.86(9.94)&0.7(1.14)&2.79(3.26)\\
				
				\cline{2-8}
				&t(2)&0.3(0.16)&4.75(0.66)&3.19(4.68)&12.78(12.71)&1.55(1.46)&3.78(3.64)\\
				
				\hline
				$\tau=0.7$&Normal&0.25(0.11)&3.40(0.59)&0.80(0.22)&2.63(0.39)&0.30(0.13)&1.12(0.29)\\
				
				\cline{2-8}
				&NormalMix&0.39(0.17)&4.77(0.76)&1.35(0.49)&4.76(0.97)&0.94(1.08)&1.85(0.77)\\
				
				\cline{2-8}
				&Laplace&0.25(0.11)&4.14(0.70)&0.88(0.25)&3.57(0.64)&0.43(0.53)&1.30(0.50)\\
				
				\cline{2-8}
				&Lognormal&0.58(0.23)&6.55(1.35)&5.32(22.78)&9.11(11.68)&1.26(1.18)&2.15(2.68)\\
				
				\cline{2-8}
				&t(2)&0.49(0.25)&6.08(0.99)&5.98(9.18)&18.73(22.33)&3.2(3.84)&4.94(5.77)\\
				
				\hline
			\end{tabular}
		}
	\end{center}
	\centering
\end{table}

\begin{table} [H]
	\def\arraystretch{1.5}
	\begin{center}
		\caption{Empirical 95$\%$ coverage probabilities under simulated gene expression data with i.i.d. $t$(2) error based on 200 replicates. \\  }\label{tab:ec}
		
		\centering
		\fontsize{10}{10}\selectfont{
			\begin{tabular}{c c l c c c c c }
				\hline
				$t(2)$ error&&  BQRVCSS& BQRVC & BVCSS & BVC \\
				\hline
				$\tau=0.3$&$\gamma_{0}(v)$ &0.800&0.875&0.570&0.630\\
				
				\cline{2-6}
				&$\gamma_{1}(v)$& 0.865&0.020&0.815&0.055\\
				
				\cline{2-6}
				&$\gamma_{2}(v)$&0.950&0.780&0.745&0.825\\
				
				\cline{2-6}
				&$\gamma_{3}(v)$&0.860&0.000&0.760&0.055\\

				\hline
				$\tau=0.5$&$\gamma_{0}(v)$&0.875&0.935&0.885&0.865\\
				
				\cline{2-6}
				&$\gamma_{1}(v)$&0.930&0.020&0.850&0.050\\
				
				\cline{2-6}
				&$\gamma_{2}(v)$&0.960&0.845&0.810&0.835\\
				
				\cline{2-6}
				&$\gamma_{3}(v)$&0.905&0.015&0.790&0.050\\

				\hline
				$\tau=0.7$&$\gamma_{0}(v)$&0.820&0.905&0.665&0.710\\
				
				\cline{2-6}
				&$\gamma_{1}(v)$&0.930&0.045&0.870&0.080\\
				
				\cline{2-6}
				&$\gamma_{2}(v)$&0.940&0.830&0.735&0.850\\
				
				\cline{2-6}
				&$\gamma_{3}(v)$&0.910&0.020&0.815&0.070\\

				\hline
			\end{tabular}
		}
	\end{center}
	\centering
\end{table}

\subsection{The estimated quantile varying coefficient functions}

\begin{figure}[H]
	\centering
	\includegraphics[width=1\textwidth]{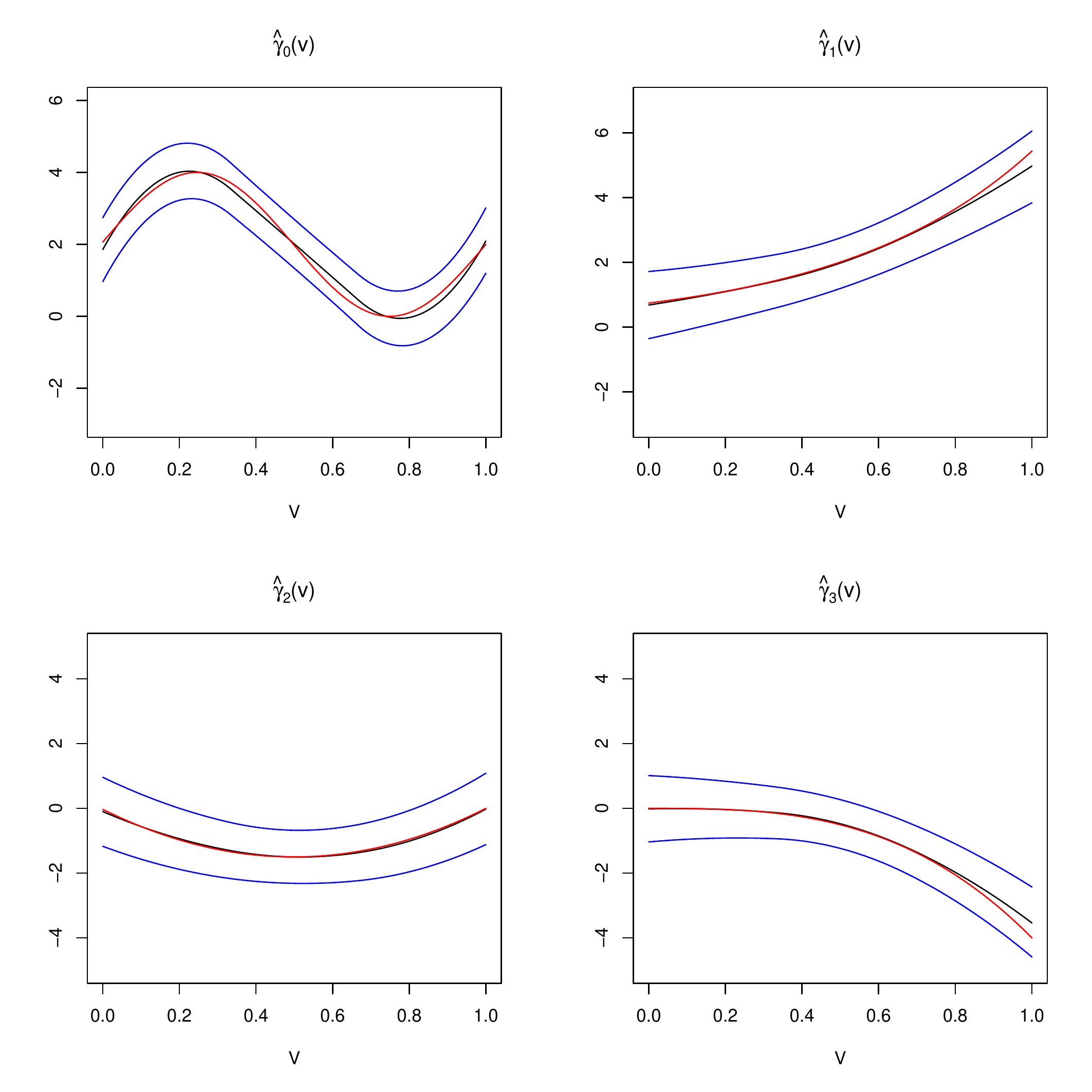}
	\caption{ Estimation of non-zero varying coefficients under the normal mixture error (Error 2)  for the proposed method (BQRVCSS) at 50$\%$ quantile level. Red line: true
		varying coefficients. Black line: posterior median estimates of varying coefficients from BQRVCSS. Blue
		lines: 95\% credible intervals for the estimated varying coefficients.	\label{fig:vc}}
\end{figure}

\subsection{Evaluation on the convergence of MCMC chains}

\begin{figure}[H]
	\centering
	\includegraphics[width=1\textwidth]{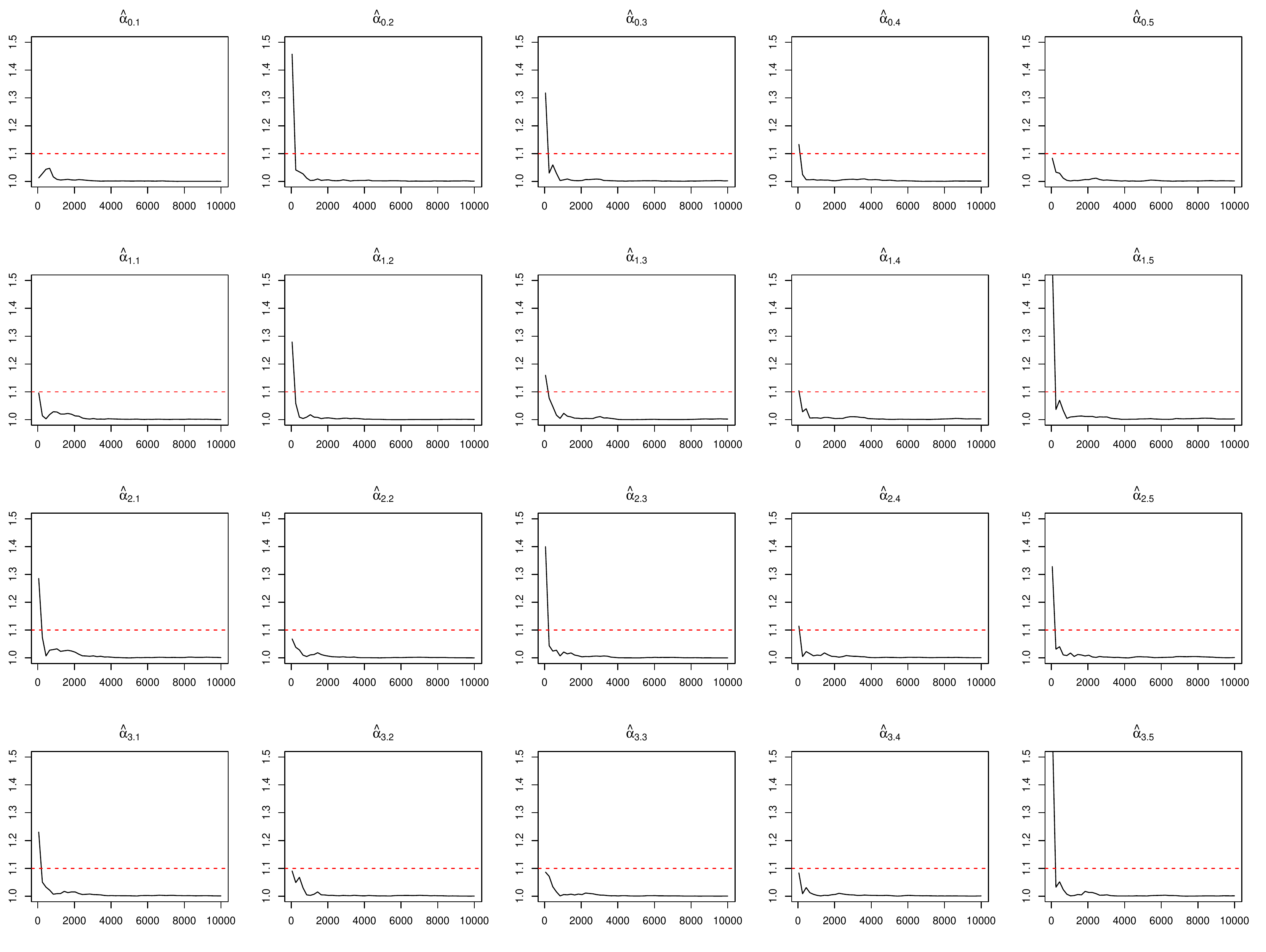}
	\caption{Potential scale reduction factor (PSRF) versus iterations for the varying functions in Figure \ref{fig:vc}.  Black line: PSRF.
		Red line: the threshold of 1.1.  $\hat{\alpha}_{j1}$ to $\hat{\alpha}_{j5} (j=0,...,3)$ denote the five estimated spline coefficients for the varying 
		coefficient function $\gamma_j$, respectively.	\label{fig:PSRF}}
\end{figure}

\subsection{Hyper-parameters sensitivity analysis}
\begin{table}[H]
		\def\arraystretch{1.5}
		
	\begin{center}
		\caption{Sensitivity analysis on the choice of the hyperparameter for $\pi_0$ using different Beta priors for the Laplace error distribution for the 30\% quantile. TIMSE: total integrated mean square error. \\ }\label{tab:4-9}
		
		\centering
		\fontsize{10}{10}\selectfont{
			\begin{tabular}{l c c c c }
				\hline
				&C&O&U&TIMSE\\
				\hline
				Beta(0.5,0.5)&0.90&0.10&0.00&0.27(0.12)\\
				Beta(1,1)&0.90&0.10&0.00&0.28(0.12)\\
				Beta(2,2)&0.90&0.10&0.00&0.28(0.11)\\
				Beta(1,5)&0.90&0.10&0.00&0.27(0.11)\\
				Beta(5,1)&0.90&0.10&0.00&0.27(0.11)\\
				\hline
			\end{tabular}
		}
	\end{center}
\centering
\end{table}
				
\begin{table}[H]
	\def\arraystretch{1.5}
	\begin{center}
		\caption{Sensitivity analysis on the choice of the hyperparameter for $\eta^2$ using different Gamma priors for the Laplace error distribution for the 30\% quantile. TIMSE: total integrated mean square error. \\ }\label{tab:00}
		
		\centering
		\fontsize{10}{10}\selectfont{
			\begin{tabular}{l c c c c }
				\hline
				&C&O&U&TIMSE\\
				\hline
				Gamma(0.1,1)&0.90&0.10&0.00&0.29(0.17)\\
				Gamma(1,1)&0.90&0.10&0.00&0.29(0.16)\\
				Gamma(1,5)&0.90&0.10&0.00&0.30(0.16)\\
				Gamma(2,5)&0.88&0.12&0.00&0.30(0.16)\\
				Gamma(5,1)&0.90&0.10&0.00&0.29(0.16)\\
				\hline
			\end{tabular}
		}
	\end{center}
	\centering
\end{table}

\begin{table}[H]
	\def\arraystretch{1.5}
	\begin{center}
		\caption{Sensitivity analysis on the choice of the hyperparameter for $\pi_0$ using different Beta priors for the Laplace error distribution for the 50\% quantile. TIMSE: total integrated mean square error.\\ }\label{tab:01}
		
		\centering
		\fontsize{10}{10}\selectfont{
			\begin{tabular}{l c c c c  }
				\hline
				&C&O&U&TIMSE\\
				\hline
				Beta(0.5,0.5)&0.92&0.08&0.00&0.22(0.05)\\
				Beta(1,1)&0.94&0.06&0.00&0.22(0.06)\\
				Beta(2,2)&0.94&0.06&0.00&0.22(0.06)\\
				Beta(1,5)&0.94&0.06&0.00&0.22(0.06)\\
				Beta(5,1)&0.92&0.08&0.00&0.22(0.06)\\
				\hline
			\end{tabular}
		}
	\end{center}
	\centering
\end{table}

\begin{table}[H]
	\def\arraystretch{1.5}
	\begin{center}
		\caption{Sensitivity analysis on the choice of the hyperparameter for $\eta^2$ using different Gamma priors for the Laplace error distribution for the 50\% quantile. TIMSE: total integrated mean square error.\\ }\label{tab:02}
		
		\centering
		\fontsize{10}{10}\selectfont{
			\begin{tabular}{l c c c c }
				\hline
				&C&O&U&TIMSE\\
				\hline
				Gamma(0.1,1)&0.96&0.04&0.00&0.22(0.05)\\
				Gamma(1,1)&0.94&0.06&0.00&0.22(0.05)\\
				Gamma(1,5)&0.94&0.06&0.00&0.23(0.05)\\
				Gamma(2,5)&0.94&0.06&0.00&0.22(0.06)\\
				Gamma(5,1)&0.94&0.06&0.00&0.22(0.05)\\
				\hline
			\end{tabular}
		}
	\end{center}
	\centering
\end{table}

\subsection{Selection of tuning parameters for frequentist methods}

We have chosen the tuning parameters for VC-adp and QRVC-adp in terms of the Schwarz-type Information Criterion (SIC):
\begin{equation*}\label{equr:sic1}
\text{SIC}(\lambda)=\text{log}\sum_{i=1}^{n}L(Y_i-\boldsymbol E_i^\top \hat{\boldsymbol\beta}-\boldsymbol Z_i^\top \hat{\boldsymbol\alpha})+\frac{\text{log}n}{2n}edf,
\end{equation*}
where $edf$ is the effective degree of freedom. For QRVC-adp, $L(\cdot)$ is the quantile check loss function, and $edf$ is the number of zero residuals which has been extensively used as a metric indicating the effective dimension of the fitted quantile regression models. Such a SIC criterion has been commonly adopted in published work on regularized quantile varying coefficient models (\cite{TangWang2012,Tang13,Noh2012}). For VC-adp, $L(\cdot)$ is the least square loss function, and $edf$ is the total number of nonzero varying coefficients (\cite{TangWang2012,wang2009shrinkage}). The R codes of VC-adp and QRVC-adp can be obtained through minor modifications to the R codes for methods proposed in  \cite{TangWang2012} available at Dr. Huixia Wang's website (\url{https://blogs.gwu.edu/judywang/software/}).  

We have examined the estimation performance of the two frequentist methods when the tuning parameters are selected using validation. Specifically, after the regularized estimates have been obtained using the training data, the prediction in terms of the check loss for QRVC-adp and least square loss for VC-adp are assessed on an independently generated testing data. For each tuning parameter across the sequence, the prediction performance is assessed on the same testing data. Therefore, the optimal tuning is corresponding to the smallest testing error. Such a method of choosing the tuning parameters is feasible in simulation as the data generating model is available, which is computationally less intensive compared to cross-validation. For illustration purpose, we have conducted the simulation under the 1st setting where gene expression data have been generated with i.i.d errors. The estimation results in Table \ref{tab:tun}	below are very close to the ones obtained in Table \ref{tab:s1} from the main text. 
		
			\begin{table} [H]
	\def\arraystretch{1.5}
	\begin{center}
		\caption{Selecting tuning parameters based on validation: estimation results in terms of total integrated mean square error (TIMSE) for simulated gene expression data with i.i.d. errors based on 100 replicates. \\  }\label{tab:tun}
		
		\centering
		\fontsize{10}{10}\selectfont{
			\begin{tabular}{ c l c c c c c c}
				\hline
				&  \multicolumn{2}{c}{$\tau=0.3$} & \multicolumn{2}{c}{$\tau=0.5$}&\multicolumn{2}{c}{$\tau=0.7$}\\
				\cline{2-7}
				&  QRVC-adp & VC-adp & QRVC-adp & VC-adp & QRVC-adp & VC-adp\\
				\hline
				Normal&0.29(0.09)&0.84(0.26)&0.28(0.13)&0.94(0.22)&0.31(0.09)&1.02(0.33)\\
				
				\cline{1-7}
				NormalMix&0.63(0.52)&1.31(0.52)&0.45(0.24)&1.05(0.28)&0.51(0.23)&1.49(0.26)\\
				
				\cline{1-7}
				Laplace&0.37(0.21)&0.96(0.17)&0.30(0.13)&1.00(0.31)&0.35(0.16)&1.23(0.25)\\
				
				\cline{1-7}
				Lognormal&0.28(0.48)&2.63(0.77)&0.51(0.65)&2.13(0.90)&0.98(0.57)&1.77(0.71)\\
				
				\cline{1-7}
				t(2)&1.19(0.91)&2.61(1.32)&0.82(0.68)&2.23(1.45)&1.18(1.13)&2.56(1.27)\\
				
				\hline
			\end{tabular}
		}
	\end{center}
	\centering
\end{table}

\subsection{Additional simulation under more complicated varying coefficient functions}

\begin{table} [H]
	\def\arraystretch{1.5}
	\begin{center}
		\caption{  Additional simulation under more complicated varying coefficient functions ($\gamma_{0}^{\star}(v)=2+2\text{sin}(6\pi v)$): estimation results in terms of total integrated mean square error (TIMSE) for simulated gene expression data with i.i.d. errors based on 100 replicates. \\  }\label{tab:apendixss}
		
		\centering
		\fontsize{10}{10}\selectfont{
			\begin{tabular}{c c l c c c c c c}
				\hline
				$\tau$&&  BQRVCSS& BQRVC & BVCSS & BVC & QRVC-adp & VC-adp\\
				\hline
				$\tau=0.3$&Normal&2.22(0.28)&4.91(0.60)&2.14(0.29)&4.20(0.47)&2.52(0.43)&2.58(0.33)\\
				
				\cline{2-8}
				&NormalMix&2.49(0.52)&5.71(0.92)&2.49(0.47)&5.25(0.72)&2.89(0.71)&2.80(0.44)\\
				
				\cline{2-8}
				&Laplace&2.42(0.56)&5.41(0.76)&2.18(0.38)&4.47(0.58)&2.90(0.70)&2.57(0.37)\\
				
				\cline{2-8}
				&Lognormal&2.17(0.41)&5.34(0.87)&3.79(1.17)&7.45(2.38)&2.85(0.79)&3.80(0.77)\\
				
				\cline{2-8}
				&t(2)&2.74(0.61)&6.67(1.69)&4.41(3.97)&9.76(5.29)&4.73(3.27)&4.32(2.60)\\
				
				\hline
				$\tau=0.5$&Normal&2.02(0.34)&4.99(0.65)&1.81(0.21)&3.83(0.39)&2.31(0.37)&2.44(0.85)\\
				
				\cline{2-8}
				&NormalMix&2.26(0.48)&5.76(0.83)&2.11(0.36)&4.87(0.61)&2.81(0.70)&2.76(1.08)\\
				
				\cline{2-8}
				&Laplace&2.21(0.50)&5.36(0.59)&1.96(0.43)&4.40(0.60)&2.68(0.72)&2.55(0.88)\\
				
				\cline{2-8}
				&Lognormal&2.34(0.48)&6.08(0.77)&3.23(1.21)&7.32(3.39)&3.07(0.87)&3.40(1.14)\\
				
				\cline{2-8}
				&t(2)&2.53(0.74)&6.16(0.89)&5.12(9.20)&9.88(7.15)&4.03(2.94)&4.82(2.91)\\
				
				\hline
				$\tau=0.7$&Normal&2.20(0.32)&5.03(0.58)&2.28(0.22)&4.05(0.41)&2.67(0.57)&3.39(1.60)\\
				
				\cline{2-8}
				&NormalMix&2.44(0.52)&5.75(0.74)&2.57(0.50)&5.34(0.82)&2.93(0.77)&4.51(2.00)\\
				
				\cline{2-8}
				&Laplace&2.42(0.39)&5.43(0.70)&2.14(0.40)&4.44(0.61)&2.85(0.52)&3.86(1.90)\\
				
				\cline{2-8}
				&Lognormal&3.07(0.92)&7.30(1.45)&3.02(1.66)&6.73(3.14)&4.22(1.56)&3.58(1.72)\\
				
				\cline{2-8}
				&t(2)&2.73(0.73)&6.68(1.11)&4.92(4.32)&9.14(3.79)&4.09(2.67)&6.16(3.66)\\
				
				\hline
			\end{tabular}
		}
	\end{center}
	\centering
\end{table}

\begin{figure}[H]
	\centering
	\includegraphics[width=1\textwidth]{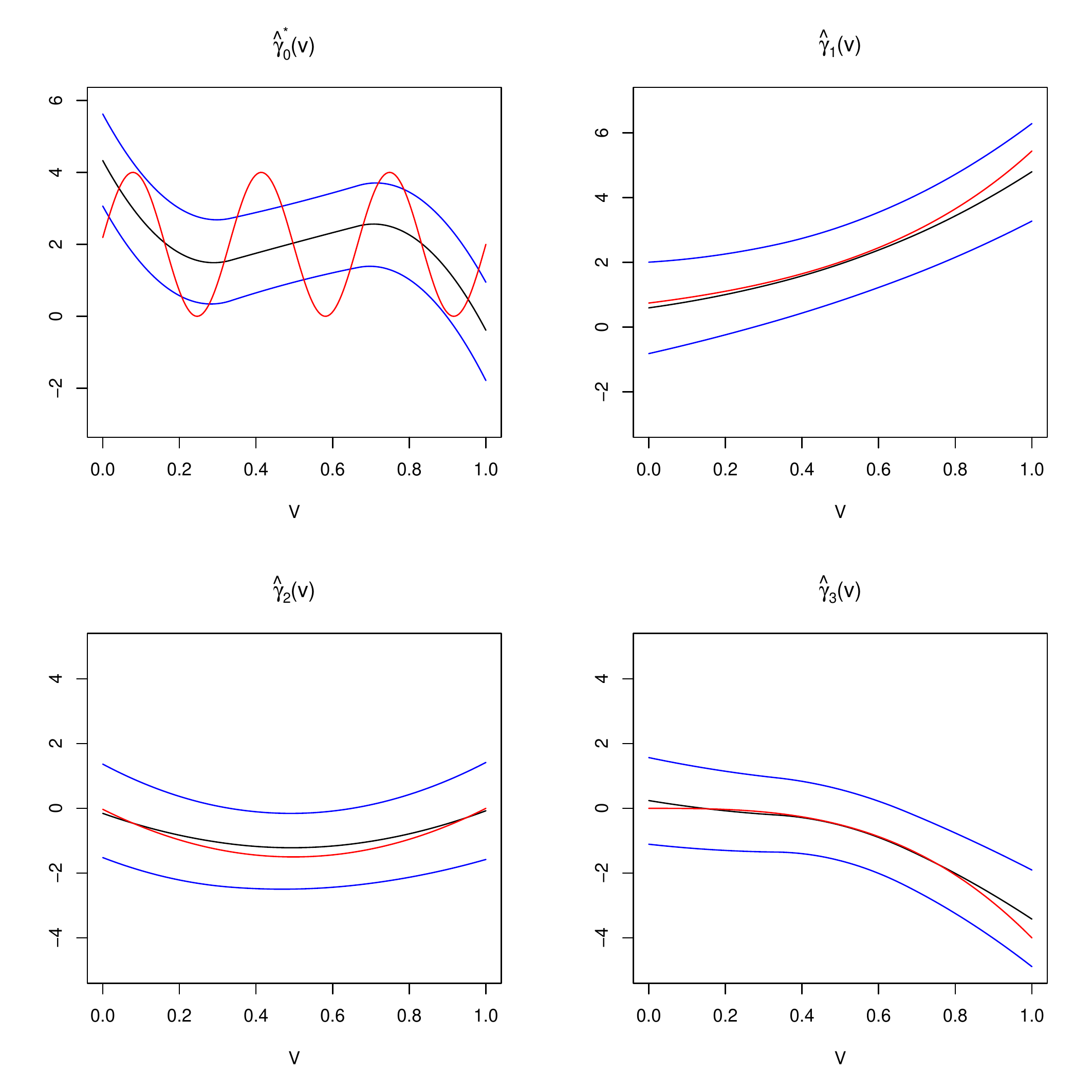}
	\caption{ Estimation of more complicated non-zero varying coefficients ($\gamma_{0}^{\star}(v)=2+2\text{sin}(6\pi v)$) under the normal mixture error (Error 2)  for the proposed method (BQRVCSS) at 50$\%$ quantile level. Red line: true varying coefficients. Black line: posterior median estimates of varying coefficients from BQRVCSS. Blue
		lines: 95\% credible intervals for the estimated varying coefficients.	\label{fig:vcs}}
\end{figure}

\begin{figure}[H]
	\centering
	\includegraphics[width=1\textwidth]{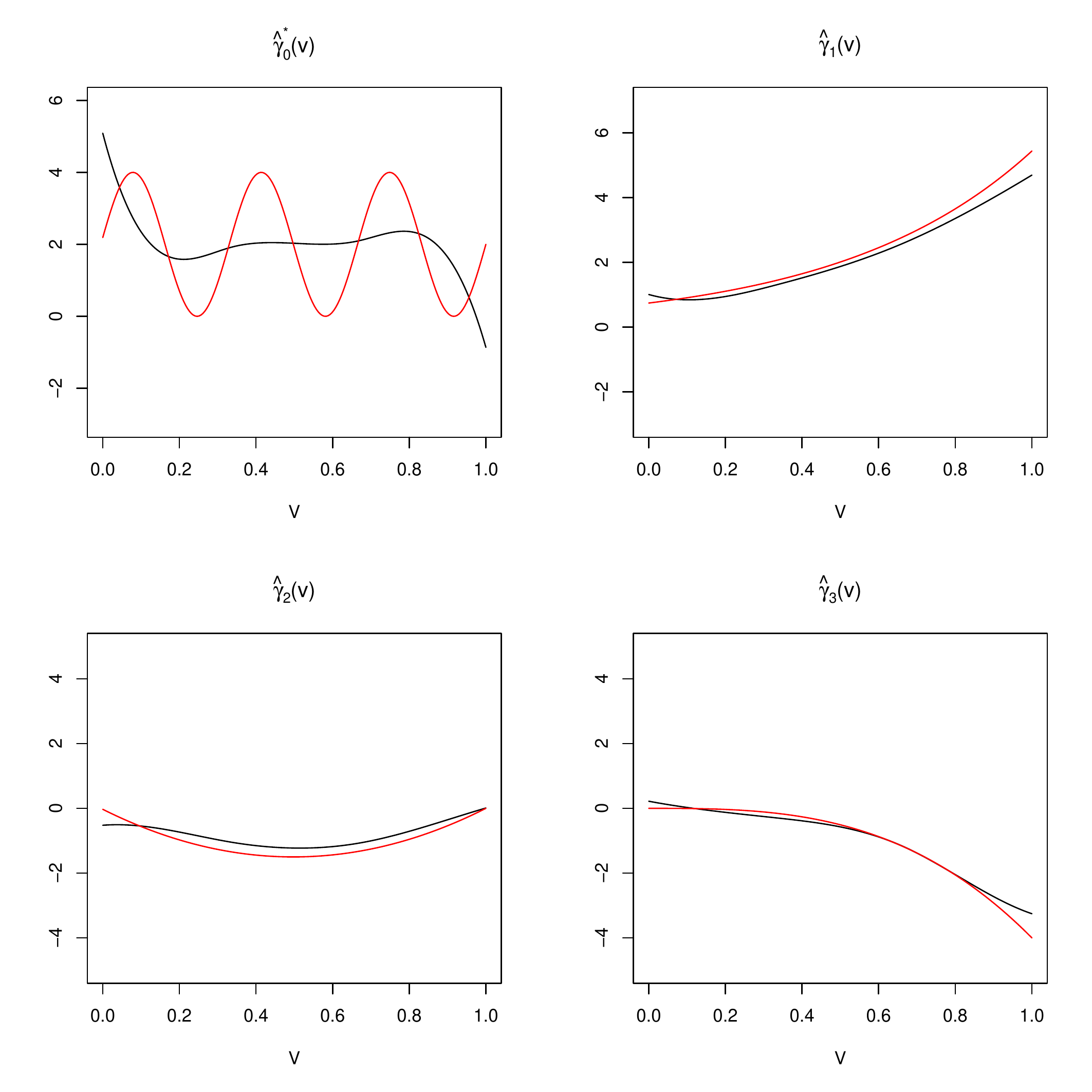}
	\caption{ Estimation of more complicated non-zero varying coefficients ($\gamma_{0}^{\star}(v)=2+2\text{sin}(6\pi v)$) under the normal mixture error (Error 2)  for the QRVC-adp at 50$\%$ quantile level. Red line: true varying coefficients. Black line: estimated varying coefficients from QRVC-adp. The confidence intervals are not available for frequentist regularized quantile varying coefficients. 	\label{fig:qrvcs}}
\end{figure}

\begin{figure}[H]
	\centering
	\includegraphics[width=1\textwidth]{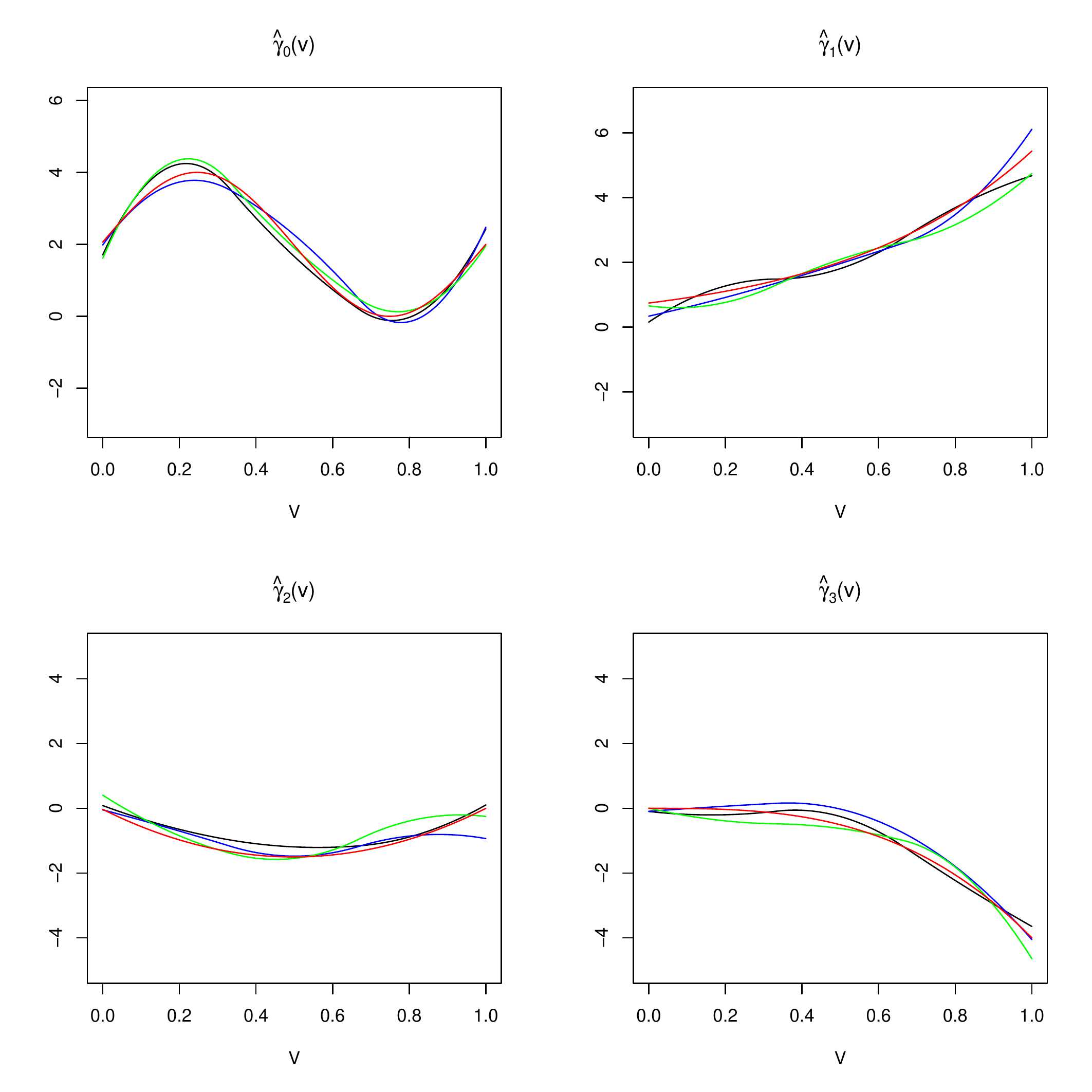}
	\caption{ Estimation of non-zero varying coefficients under the normal mixture error (Error 2)  for the proposed method (BQRVCSS) at 50$\%$ quantile level. Red line: true
		varying coefficients. Black, Blue and Green lines: posterior median estimates of varying coefficients from BQRVCSS under 25$\%$, 50$\%$ and 75$\%$ IMSE respectively.	\label{fig:vcmse}}
\end{figure}

\section{Sensitivity analysis on smoothness specification}
\label{Appendix:Key1}
Let O denote the degree of B spline basis and $N_n$ denote the number of interior knots. For quadratic and cubic splines corresponding to O=2 and O=3 respectively, we conduct a sensitivity analysis for the proposed model. 
\begin{table} [H]
	\def\arraystretch{1.5}
	\begin{center}
		\caption{Sensitivity analysis on smoothness specification for the Laplace error distribution for the 30\% quantile. TIMSE: total integrated mean square error.\\ }\label{tab:4-10}
		
		\centering
		\fontsize{10}{10}\selectfont{
			\begin{tabular}{l l c c c c c }
				\hline
				O=2&$N_n$& 1& 2 & 3 & 4 & 5 \\
				\hline
				Laplace&C&0.88&0.90&0.92&0.89&0.91\\
				&O&0.12&0.10&0.08&0.11&0.09\\
				&U&0.00&0.00&0.00&0.00&0.00\\
				&TIMSE&0.33(0.19)&0.28(0.12)&0.31(0.14)&0.24(0.12)&0.25(0.15)\\

				\hline
					O=3&$N_n$& 1& 2 & 3 & 4 & 5 \\
				\hline
				Laplace&C&0.89&0.90&0.92&0.86&0.88\\
				&O&0.11&0.10&0.08&0.14&0.12\\
				&U&0.00&0.00&0.00&0.00&0.00\\
				&TIMSE&0.25(0.11)&0.28(0.12)&0.28(0.15)&0.26(0.19)&0.25(0.16)\\

				\hline
			\end{tabular}
		}
	\end{center}
	\centering
\end{table}

\begin{table} [H]
	\def\arraystretch{1.5}
	\begin{center}
		\caption{Sensitivity analysis on smoothness specification for the Normal error distribution for the 30\% quantile. TIMSE: total integrated mean square error.\\ }\label{tab:4-11}
		
		\centering
		\fontsize{10}{10}\selectfont{
			\begin{tabular}{l l c c c c c }
				\hline
				O=2&$N_n$& 1& 2 & 3 & 4 & 5 \\
				\hline
				Normal&C&0.97&0.96&0.98&0.95&0.94\\
				&O&0.03&0.04&0.04&0.05&0.06\\
				&U&0.00&0.00&0.00&0.00&0.00\\
				&TIMSE&0.26(0.12)&0.22(0.09)&0.29(0.16)&0.23(0.12)&0.22(0.18)\\
				
				\hline
				O=3&$N_n$& 1& 2 & 3 & 4 & 5 \\
				\hline
				Normal&C&0.96&0.94&0.97&0.94&0.95\\
				&O&0.04&0.06&0.03&0.06&0.05\\
				&U&0.00&0.00&0.00&0.00&0.00\\
				&TIMSE&0.24(0.09)&0.26(0.14)&0.21(0.10)&0.25(0.19)&0.24(0.12)\\
				
				\hline
			\end{tabular}
		}
	\end{center}
	\centering
\end{table}

\begin{table} [H]
	\def\arraystretch{1.5}
	\begin{center}
		\caption{Sensitivity analysis on smoothness specification for the Laplace error distribution for the 50\% quantile. TIMSE: total integrated mean square error.\\ }\label{tab:4-12}
		
		\centering
		\fontsize{10}{10}\selectfont{
			\begin{tabular}{l l c c c c c }
				\hline
				O=2&$N_n$& 1& 2 & 3 & 4 & 5 \\
				\hline
				Laplace&C&0.96&0.94&0.92&0.95&0.96\\
				&O&0.04&0.06&0.08&0.05&0.04\\
				&U&0.00&0.00&0.00&0.00&0.00\\
				&TIMSE&0.25(0.11)&0.21(0.09)&0.29(0.16)&0.28(0.11)&0.25(0.19)\\
				
				\hline
				O=3&$N_n$& 1& 2 & 3 & 4 & 5 \\
				\hline
				Laplace&C&0.95&0.93&0.94&0.96&0.93\\
				&O&0.05&0.07&0.06&0.04&0.07\\
				&U&0.00&0.00&0.00&0.00&0.00\\
				&TIMSE&0.24(0.07)&0.31(0.14)&0.26(0.12)&0.22(0.16)&0.26(0.13)\\
				
				\hline
			\end{tabular}
		}
	\end{center}
	\centering
\end{table}

\begin{table} [H]
	\def\arraystretch{1.5}
	\begin{center}
		\caption{Sensitivity analysis on smoothness specification for the Normal error distribution for the 50\% quantile. TIMSE: total integrated mean square error.\\ }\label{tab:4-13}
		
		\centering
		\fontsize{10}{10}\selectfont{
			\begin{tabular}{l l c c c c c }
				\hline
				O=2&$N_n$& 1& 2 & 3 & 4 & 5 \\
				\hline
				Normal&C&0.97&0.98&0.96&0.99&0.98\\
				&O&0.03&0.02&0.04&0.01&0.02\\
				&U&0.00&0.00&0.00&0.00&0.00\\
				&TIMSE&0.21(0.06)&0.23(0.13)&0.22(0.07)&0.24(0.14)&0.22(0.09)\\

				\hline
				O=3&$N_n$& 1& 2 & 3 & 4 & 5 \\
				\hline
				Normal&C&0.98&0.96&0.98&0.98&0.97\\
				&O&0.02&0.04&0.02&0.02&0.03\\
				&U&0.00&0.00&0.00&0.00&0.00\\
				&TIMSE&0.19(0.07)&0.29(0.11)&0.25(0.07)&0.24(0.14)&0.23(0.08)\\
				
				\hline
			\end{tabular}
		}
	\end{center}
	\centering
\end{table}

\section{Posterior inference}
\label{Appendix:Key3}

\subsection{Posterior inference for BQRVCSS}

\subsubsection{Bayesian hierarchical model}

\begin{equation*}
Y_i=\sum_{k=1}^{q} E_{ik} \beta_k +\sum_{j=0}^{p}\boldsymbol \alpha_j ^\top\boldsymbol Z_{ij}+\kappa_1\tilde{u}_i+\theta^{-\frac{1}{2}}\kappa_2\sqrt{\tilde{u}_i}W_i,
\end{equation*} 

\begin{equation*} 
\tilde{u_1},...,\tilde{u_n} \thicksim
\prod_{i=1}^{n} \theta \text{exp}(-\theta \tilde{u_i}), i=1,...,n,
\end{equation*}
\begin{equation*}
W_1,...,W_n \thicksim \prod_{i=1}^{n} \frac{1}{\sqrt{2 \pi}} \text{exp}(-\frac{1}{2}W_i^2) , i=1,...,n,
\end{equation*}
\begin{equation*}
\boldsymbol\alpha_j|g_j \text{ } \overset{\mathrm{ind}}{\thicksim} \text{ } (1-\pi_0)\text{N}_d(0,g_j \textbf{I}_d)+\pi_0\delta_0(\boldsymbol\alpha_j), j=1,...,p,
\end{equation*}
\begin{equation*}
g_j|\eta^2 \text{ } \overset{\mathrm{ind}}{\thicksim} \text{ } \text{Gamma} \Big(\frac{d+1}{2},\frac{\eta^2}{2}\Big), j=1,...,p,
\end{equation*}

\begin{equation*}
\pi_0 \thicksim \text{Beta}(e,f),
\end{equation*}
\begin{equation*}
\theta \thicksim \text{Gamma}(a,b),
\end{equation*}
\begin{equation*}
\eta^2 \thicksim \text{Gamma}(c,m),
\end{equation*}

\begin{equation*}
\boldsymbol\beta \thicksim \text{N}_q(0,\boldsymbol \Sigma_{\boldsymbol\beta}),
\end{equation*}
\begin{equation*}
\boldsymbol\alpha_0 \thicksim \text{N}_d(0,\boldsymbol \Sigma_{\boldsymbol\alpha_0}).
\end{equation*}

\subsubsection{Gibbs Sampler}

$\bullet$ The full conditional distribution of $\tilde{u_i}$, $i=1,...,n,$
\begin{equation*}
\setlength{\jot}{10pt}
\begin{aligned}
p(\tilde{u_i}|&\text{rest})\\
& \propto \frac{1}{\sqrt{2\pi \theta^{-1}\kappa_2^2 \tilde{u_i}}}\text{exp}\Big({-\frac{1}{2}\frac{(\boldsymbol Y_i-\boldsymbol E_i^\top \boldsymbol \beta-\sum_{j=0}^{p}\boldsymbol \alpha_j ^\top\boldsymbol Z_{ij} - \kappa_1 \tilde{u_i})^2}{\kappa_2^2 \theta^{-1} \tilde{u_i}}})\theta \text{exp}(-\theta \tilde{u_i}\Big)\\
&\propto (\tilde{u_i})^{-\frac{1}{2}}\text{exp}\Big({-\frac{1}{2}\frac{(\boldsymbol Y_i-\boldsymbol E_i^\top \boldsymbol \beta-\sum_{j=0}^{p}\boldsymbol \alpha_j ^\top\boldsymbol Z_{ij})}{\kappa_2^2 \theta^{-1} \tilde{u_i}}}-\frac{1}{2}\frac{\kappa_1^2\tilde{u_i}}{\theta^{-1}\kappa_2^2}-\theta \tilde{u_i}\Big)\\
&\propto (\tilde{u_i})^{-\frac{1}{2}}\text{exp}\Bigg(-\frac{1}{2}\Big((\frac{\theta \kappa_1^2}{\kappa_2^2}+2\theta)\tilde{u_i}+\frac{\theta (\boldsymbol Y_i-\boldsymbol E_i^\top \boldsymbol \beta-\sum_{j=0}^{p}\boldsymbol \alpha_j ^\top\boldsymbol Z_{ij}}{\kappa_2^2}\frac{1}{\tilde{u_i}}\Big)\Bigg).
\end{aligned}
\end{equation*}
Hence, it follows that 
\begin{equation*}
\tilde{u_i}^{-1}|\text{rest}  \thicksim\text{Inverse-Gaussian}(\sqrt{\frac{\kappa_1^2+2\kappa_2^2}{(Y_i-\boldsymbol E_i^\top \boldsymbol\beta-Z_i^\top \boldsymbol\alpha)^2}},(\frac{\theta \kappa_1^2}{\kappa_2^2}+2\theta)).
\end{equation*}

$\bullet$ The full conditional distribution of $\boldsymbol\alpha_j$, $j=1,...,p,$
\begin{equation*}
\setlength{\jot}{10pt}
\begin{aligned}
p(\boldsymbol\alpha_j|&\text{rest}) \\
& \propto \prod_{i=1}^{n}\text{exp}\Big(-\frac{\theta}{2\kappa_2^2 \tilde{u_i}}(\boldsymbol Y_i-\boldsymbol Z_{i,-j}^{\top}\boldsymbol\alpha_{-j}-\boldsymbol Z_{ij}^\top \boldsymbol\alpha_j-\boldsymbol E_i^\top \boldsymbol\beta-\kappa_1 \tilde{u_i} )^2\Big)\\
&\times \Bigg((1-\pi_0)(2\pi g_j)^{-\frac{d}{2}}\text{exp}\Big(-\frac{1}{2}\boldsymbol\alpha_j^\top(g_j \textbf{I}_d)^{-1}\boldsymbol\alpha_j \Big)\textbf{I}_{(\boldsymbol\alpha_j \neq 0)}+\pi_0 \delta_0 (\boldsymbol\alpha_j)\Bigg) .
\end{aligned}
\end{equation*}

Let $l_j=p(\boldsymbol\alpha_j=0|\text{rest})$, then the full  conditional posterior distribution of $\boldsymbol\alpha_j(j=1,...,p)$ is given as:
\begin{equation*}
\boldsymbol\alpha_j|\text{rest} \thicksim (1-l_j)\text{N}_d(\boldsymbol\mu_j,\boldsymbol\Sigma_j) + l_j\delta_0(\boldsymbol\alpha_j),
\end{equation*}
where 
\begin{equation*}
\boldsymbol\mu_j=\boldsymbol\Sigma_j\theta \kappa_2^{-2}\sum_{i=1}^{n}\frac{\boldsymbol Z_{ij}}{\tilde{u_i}}(Y_i-\boldsymbol Z_{i,-j}^\top\boldsymbol \alpha_{-j} -\boldsymbol E_i^\top \boldsymbol\beta-\kappa_1 \tilde{u_i}),
\end{equation*}
\begin{equation*}
\boldsymbol\Sigma_j=(\theta \kappa_2^{-2}\sum_{i=1}^{n}\frac{1}{\tilde{u_i}}\boldsymbol Z_{ij}\boldsymbol Z_{ij}^\top+g_j^{-1}\textbf{I}_d)^{-1},
\end{equation*}

and 
\begin{equation*}
l_j=\frac{\pi_0}{\pi_0+(1-\pi_0)|g_j \textbf{I}_d|^{-\frac{1}{2}}|\boldsymbol\Sigma_j |^{\frac{1}{2}}\text{exp}(\frac{1}{2}\boldsymbol\mu_j^\top\boldsymbol\Sigma_j \boldsymbol\mu_j )}.
\end{equation*}

Hence, the posterior distribution of $\boldsymbol\alpha_j$ is a mixture of a multivariate normal distribution and a point mass at 0.  At each iteration of MCMC, $\boldsymbol\alpha_j$ is drawn from $\text{N}_d(\boldsymbol\mu_j,\boldsymbol\Sigma_j)$ with probability $(1-l_j)$ and is set to 0 with probability $l_j$.

$\bullet$ The full conditional distribution of $\theta$ is
\begin{equation*}
\setlength{\jot}{10pt}
\begin{aligned}
p(\theta|&\text{rest})\\
& \propto  \prod_{i=1}^{n}\sqrt{\theta}\text{exp}\Big(-\frac{\theta(\boldsymbol Y_i-\sum_{j=0}^{p}\boldsymbol \alpha_j ^\top\boldsymbol Z_{ij}-\boldsymbol E_i^\top \boldsymbol\beta-\kappa_1 \tilde{u_i}) ^2}{\kappa_2^2\tilde{u_i}} \Big)\times \prod_{i=1}^{n}[\theta \text{exp}(-\theta \tilde{u_i})] \\
& \times \theta^{a-1}\text{exp}(-b\theta) \\
&\propto \theta^{\frac{3}{2}n+a-1}\text{exp}\Bigg(-\Big(\frac{1}{2}\sum_{i=1}^{n}\frac{(\boldsymbol Y_i-\sum_{j=0}^{p}\boldsymbol \alpha_j ^\top\boldsymbol Z_{ij}-\boldsymbol E_i^\top \boldsymbol\beta-\kappa_1 \tilde{u_i}) ^2}{\kappa_2^2\tilde{u_i}} +\sum_{i=1}^{n}\tilde{u_i}+b \Big)\theta \Bigg).
\end{aligned}
\end{equation*}
Therefore, 
\begin{equation*}
\theta|\text{rest} \thicksim \text{Gamma}\Big(\frac{3}{2}n+a, \frac{1}{2}\sum_{i=1}^{n}\frac{(\boldsymbol Y_i-\sum_{j=0}^{p}\boldsymbol \alpha_j ^\top\boldsymbol Z_{ij}-\boldsymbol E_i^\top \boldsymbol\beta-\kappa_1 \tilde{u_i}) ^2}{\kappa_2^2\tilde{u_i}} +\sum_{i=1}^{n}\tilde{u_i}+b  \Big).
\end{equation*}	
$\bullet$ The full conditional distribution of $\eta^2$ is
\begin{equation*}
\setlength{\jot}{10pt}
\begin{aligned}
p(\eta^2|\text{rest}) 
& \propto  \prod_{j=1}^{p}[(\frac{\eta^{2}}{2})^{\frac{d+1}{2}}\text{exp}(-\frac{\eta^{2}}{2}g_j )]\times (\eta^2)^{c-1} \text{exp}(-m \eta^2)\\
&\propto (\eta^{2})^{\frac{(d+1)p}{2}+c-1}\text{exp}(-(\frac{1}{2}\sum_{j=1}^{p}g_j+m )\eta^2).
\end{aligned}
\end{equation*}
It follows that  
\begin{equation*}
\eta^2|\text{rest}\thicksim \text{Gamma}\Big(\frac{(d+1)p}{2}+c, \frac{1}{2}\sum_{j=1}^{p}g_j+m\Big).
\end{equation*}

$\bullet$ The full conditional distribution of $g_j$ $ (j=1,...,p)$ is
\begin{equation*}
\setlength{\jot}{10pt}
\begin{aligned}
p(g_j|&\text{rest}) \\
& \propto  \Bigg((1-\pi_0)(2\pi g_j)^{-\frac{d}{2}}\text{exp}\Big(-\frac{1}{2}\boldsymbol\alpha_j^\top(g_j\textbf{I}_d)^{-1}\boldsymbol\alpha_j \Big)\textbf{I}_{(\boldsymbol\alpha_j \neq 0)}+\pi_0 \delta_0 (\boldsymbol\alpha_j) \Bigg)\times g_j^{\frac{d-1}{2}}\text{exp}(-\frac{\eta^2}{2}g_j).
\end{aligned}
\end{equation*}
Then,
\begin{equation*}
g_j^{-1}|\text{rest} \thicksim \begin{cases}
\scalebox{1}{Inverse-Gamma($\frac{d+1}{2}$,\, $\frac{\eta^{2}}{2}$)}& { \text{if} \; \boldsymbol\alpha_{j} = 0} \\[6pt]
\scalebox{1}{Inverse-Gaussian($\sqrt{\frac{\eta^2}{\boldsymbol\alpha_j^\top \boldsymbol\alpha_j}}$,$\eta^{2}$)}& { \text{if} \; \boldsymbol\alpha_{j} \neq 0}
\end{cases}.
\end{equation*}

$\bullet$ The full conditional distribution of $\pi_0$ is
\begin{equation*}
\setlength{\jot}{10pt}
\begin{aligned}
p(\pi_0|&\text{rest}) \\
& \propto \prod_{j=1}^{p} \Bigg((1-\pi_0)(2\pi g_j)^{-\frac{d}{2}}\text{exp}\Big(-\frac{1}{2}\boldsymbol\alpha_j^\top(g_j\textbf{I}_d)^{-1}\boldsymbol\alpha_j \Big)\textbf{I}_{(\boldsymbol\alpha_j \neq 0)}+\pi_0 \delta_0 (\boldsymbol\alpha_j) \Bigg)\times \pi_0^{e-1}(1-\pi_0)^{f-1}
\end{aligned}
\end{equation*}
Let 
\begin{equation*}
Q_j=\begin{cases}
\scalebox{1}{0}& { \text{if} \; \boldsymbol\alpha_{j} = 0} \\[6pt]
\scalebox{1}{1}& { \text{if} \; \boldsymbol\alpha_{j} \neq 0}
\end{cases},
\end{equation*}
consequently,
\begin{equation*}
\pi_0|\text{rest} \propto \text{Beta}\Big(1+p-\sum_{j=1}^{p}Q_j+e,\sum_{j=1}^{p}Q_j+f\Big)
\end{equation*}

$\bullet$ The full conditional distribution of $\boldsymbol\beta$ is
\begin{equation*}
\setlength{\jot}{10pt}
\begin{aligned}
p(\boldsymbol\beta|&\text{rest}) \\
& \propto \prod_{i=1}^{n}\text{exp}\Big(-\frac{\theta}{2\kappa_2^2 \tilde{u_i}}(\boldsymbol Y_i-\sum_{j=0}^{p}\boldsymbol \alpha_j ^\top\boldsymbol Z_{ij}-\boldsymbol E_i^\top \boldsymbol\beta-\kappa_1 \tilde{u_i} )^2\Big)\text{exp}(-\frac{1}{2}\boldsymbol\beta^\top \boldsymbol\Sigma_{\boldsymbol\beta}^{-1}\boldsymbol\beta) \\
& \propto \text{exp}\Bigg(-\frac{1}{2}\Big(\boldsymbol\beta^\top(\sum_{i=1}^{n}\frac{\theta \boldsymbol E_i \boldsymbol E_i^\top}{\kappa_2^2 \tilde{u_i}}+\boldsymbol\Sigma_{\boldsymbol\beta}^{-1})\boldsymbol\beta-2\sum_{i=1}^{n}\frac{\theta}{\kappa_2^2 \tilde{u_i}}(\boldsymbol Y_i-\sum_{j=0}^{p}\boldsymbol \alpha_j ^\top\boldsymbol Z_{ij}-\kappa_1\tilde{u_i})\boldsymbol E_i^\top \boldsymbol\beta \Big) \Bigg),
\end{aligned}
\end{equation*}
therefore, we have
\begin{equation*}
\boldsymbol\beta|\text{rest} \thicksim \text{N}_q (\boldsymbol\mu_{\beta^{\star}},\boldsymbol\Sigma_{\beta^{\star}}),
\end{equation*}
with 
\begin{equation*}
\boldsymbol\Sigma_{\beta^{\star}}=(\sum_{i=1}^{n}\frac{\theta \boldsymbol E_i \boldsymbol E_i^\top}{\kappa_2^2 \tilde{u_i}}+\boldsymbol\Sigma_{\boldsymbol\beta}^{-1})^{-1},
\end{equation*}
and
\begin{equation*}
\boldsymbol\mu_{\beta^{\star}}=\Sigma_{\beta^{\star}}\Big(\sum_{i=1}^{n}\frac{\theta}{\kappa_2^2 \tilde{u_i}}(Y_i-\sum_{j=0}^{p}\boldsymbol \alpha_j ^\top\boldsymbol Z_{ij}-\kappa_1\tilde{u_i})\boldsymbol E_i^\top \Big)^\top.
\end{equation*}

$\bullet$ Similarly the full conditional distribution of $\boldsymbol\alpha_0$ is derived as
\begin{equation*}
\setlength{\jot}{10pt}
\boldsymbol\alpha_0|\text{rest} 
\thicksim \text{N}_d(\boldsymbol\mu_0,\boldsymbol\Sigma_0),
\end{equation*}
with
\begin{equation*}
\boldsymbol\Sigma_0 = (\sum_{i=1}^{n}\frac{\theta \boldsymbol Z_{i0} \boldsymbol Z_{i0}^\top}{\kappa_2^2 \tilde{u_i}}+\boldsymbol\Sigma_{\boldsymbol\alpha_0}^{-1})^{-1}
\end{equation*}
and 
\begin{equation*}
\boldsymbol\mu_0 = \boldsymbol\Sigma_0\Big(\sum_{i=1}^{n}\frac{\theta}{\kappa_2^2 \tilde{u_i}}(Y_i-\boldsymbol E_i^\top \boldsymbol\beta-\sum_{j=1}^{p}\boldsymbol \alpha_j ^\top\boldsymbol Z_{ij}-\kappa_1\tilde{u_i})\boldsymbol Z_{i0}^\top \Big)^\top.
\end{equation*}

\subsection{Posterior inference for BQRVC}

\subsubsection{Bayesian hierarchical model}

\begin{equation*}
Y_i=\boldsymbol E_i^\top \boldsymbol\beta+\sum_{j=0}^{p}\boldsymbol \alpha_j ^\top\boldsymbol Z_{ij} + \kappa_1 \tilde{u_i} + \kappa_2 \theta^{-\frac{1}{2}}\sqrt{\tilde{u_i}}W_i , i=1,...,n,
\end{equation*} 

\begin{equation*} 
\tilde{u_1},...,\tilde{u_n} \thicksim \prod_{i=1}^{n} \theta \text{exp}(-\theta \tilde{u_i}), i=1,...,n,
\end{equation*}
\begin{equation*}
W_1,...,W_n \thicksim \prod_{i=1}^{n} \frac{1}{\sqrt{2 \pi}} \text{exp}(-\frac{1}{2}W_i^2) , i=1,...,n,
\end{equation*}
\begin{equation*}
\boldsymbol\alpha_j|g_j \text{ } \overset{\mathrm{ind}}{\thicksim} \text{ } \text{N}_d(0,g_j \textbf{I}_d), j=1,...,p,
\end{equation*}
\begin{equation*}
g_j|\eta^2 \text{ } \overset{\mathrm{ind}}{\thicksim} \text{ } \text{Gamma} \Big(\frac{d+1}{2},\frac{\eta^2}{2}\Big), j=1,...,p,
\end{equation*}
\begin{equation*}
\pi_0 \thicksim \text{Beta}(e,f),
\end{equation*}
\begin{equation*}
\theta \thicksim \text{Gamma}(a,b),
\end{equation*}
\begin{equation*}
\eta^2 \thicksim \text{Gamma}(c,m),
\end{equation*}
\begin{equation*}
\boldsymbol\beta \thicksim \text{N}_q(0,\boldsymbol\Sigma_{\boldsymbol\beta}),
\end{equation*}
\begin{equation*}
\boldsymbol\alpha_0 \thicksim \text{N}_d(0,\boldsymbol\Sigma_{\boldsymbol\alpha_0}).
\end{equation*}

\subsubsection{Gibbs Sampler}
$\bullet$ The full conditional distribution of $\tilde{u_i}$, $i=1,...,n,$
\begin{equation*}
\setlength{\jot}{10pt}
\begin{aligned}
\pi(\tilde{u_i}|\text{rest}) 
& \propto \frac{1}{\sqrt{2\pi \theta^{-1}\kappa_2^2 \tilde{u_i}}}\text{exp}\Big({-\frac{1}{2}\frac{(Y_i-\boldsymbol E_i^\top \boldsymbol\beta-\boldsymbol Z_i^\top \boldsymbol\alpha - \kappa_1 \tilde{u_i})^2}{\kappa_2^2 \theta^{-1} \tilde{u_i}}} \Big)\theta \text{exp}(-\theta \tilde{u_i})\\
&\propto (\tilde{u_i})^{-\frac{1}{2}}\text{exp}\Big({-\frac{1}{2}\frac{(Y_i-\boldsymbol E_i^\top \boldsymbol\beta-\boldsymbol Z_i^\top \boldsymbol\alpha)^2}{\kappa_2^2 \theta^{-1} \tilde{u_i}}}-\frac{1}{2}\frac{\kappa_1^2\tilde{u_i}}{\theta^{-1}\kappa_2^2}-\theta \tilde{u_i}\Big)\\
&\propto (\tilde{u_i})^{-\frac{1}{2}}\text{exp}\Bigg(-\frac{1}{2}\Big((\frac{\theta \kappa_1^2}{\kappa_2^2}+2\theta)\tilde{u_i}+\frac{\theta (Y_i-\boldsymbol E_i^\top \boldsymbol\beta-Z_i^\top \boldsymbol\alpha)^2}{\kappa_2^2}\frac{1}{\tilde{u_i}}\Big)\Bigg)
\end{aligned}
\end{equation*}

Then, the full conditional distribution of $\tilde{u_i}$ is 
\begin{equation*}
\tilde{u_i}^{-1}|\text{rest}  \thicksim\text{Inverse-Gaussian}(\sqrt{\frac{\kappa_1^2+2\kappa_2^2}{(Y_i-\boldsymbol E_i^\top \boldsymbol\beta-Z_i^\top \boldsymbol\alpha)^2}},(\frac{\theta \kappa_1^2}{\kappa_2^2}+2\theta)).
\end{equation*}

$\bullet$ The full conditional distribution of $g_j (j=1,...,p)$ is
\begin{equation*}
\setlength{\jot}{10pt}
\begin{aligned}
\pi(g_j|\text{rest}) 
& \propto  (2\pi g_j)^{-\frac{d}{2}} \text{exp}\Big(-\frac{1}{2}\boldsymbol\alpha_j^\top(g_j\textbf{I}_d)^{-1}\boldsymbol\alpha_j \Big)\times g_j^{\frac{d-1}{2}}\text{exp}(-\frac{\eta^2}{2}g_j)\\
& \propto   g_j^{-\frac{1}{2}}\text{exp}\Big(-\frac{1}{2}(\eta^2 g_j+\boldsymbol\alpha_j^\top \boldsymbol\alpha_j \frac{1}{g_j})\Big)
\end{aligned}
\end{equation*}
It follows that
\begin{equation*}
g_j^{-1}|\text{rest}  \thicksim\text{Inverse-Gaussian}(\sqrt{\frac{\eta^2}{\boldsymbol\alpha_j^\top \boldsymbol\alpha_j}},\eta^2).
\end{equation*}

$\bullet$ The full conditional distribution of $\boldsymbol\alpha_j$, $j=1,...,p,$
\begin{equation*}
\setlength{\jot}{10pt}
\begin{aligned}
p(\boldsymbol\alpha_j|&\text{rest}) \\
& \propto \prod_{i=1}^{n}\text{exp}\Big(-\frac{\theta}{2\kappa_2^2 \tilde{u_i}}(Y_i-\boldsymbol Z_{i,-j}^{\top}\boldsymbol\alpha_{-j}-\boldsymbol Z_{ij}^\top \boldsymbol\alpha_j-\boldsymbol E_i^\top \boldsymbol\beta-\kappa_1 \tilde{u_i} )^2\Big)\\
&\times (2\pi g_j)^{-\frac{d}{2}}\text{exp}\Big(-\frac{1}{2}\boldsymbol\alpha_j^\top(g_j \textbf{I}_d)^{-1}\boldsymbol\alpha_j \Big) \\
& \propto \text{exp}\Big(-\frac{1}{2}\theta \kappa_2^{-2} \sum_{i=1}^{n}\frac{1}{\tilde{u_i}}(Y_i-\boldsymbol Z_{i,-j}^{\top}\boldsymbol\alpha_{-j}-\boldsymbol E_i^\top \boldsymbol\beta-\kappa_1 \tilde{u_i}) ^2\Big)\\
&\times \text{exp}\Bigg(-\frac{1}{2}\Big(\alpha_j^\top(\theta \kappa_2^{-2}\sum_{i=1}^{n}\frac{\boldsymbol Z_{ij}\boldsymbol Z_{ij}^\top}{\tilde{u_i}}+g_j^{-1}\textbf{I}_d)\boldsymbol\alpha_j-2\theta \kappa_2^{-2}\sum_{i=1}^{n}\frac{1}{\tilde{u_i}}(Y_i-\boldsymbol Z_{i,-j}^{\top}\boldsymbol\alpha_{-j}-\boldsymbol E_i^\top \boldsymbol\beta-\kappa_1 \tilde{u_i})\boldsymbol Z_{ij}^\top \boldsymbol\alpha_j \Big)\Bigg)
\end{aligned}
\end{equation*}
Denote the covariance
\begin{equation*}
\boldsymbol\Sigma_j=(\theta \kappa_2^{-2}\sum_{i=1}^{n}\frac{1}{\tilde{u_i}}\boldsymbol Z_{ij}\boldsymbol Z_{ij}^\top+g_j^{-1}\textbf{I}_d)^{-1}
\end{equation*}
and the mean
\begin{equation*}
\boldsymbol\mu_j=\boldsymbol\Sigma_j\theta \kappa_2^{-2}\sum_{i=1}^{n}\frac{\boldsymbol Z_{ij}}{\tilde{u_i}}(Y_i-\boldsymbol Z_{i,-j}^{\top}\boldsymbol\alpha_{-j}-\boldsymbol E_i^\top \boldsymbol\beta-\kappa_1 \tilde{u_i}),
\end{equation*}
then we have
\begin{equation*}
\boldsymbol\alpha_j|\text{rest} \thicksim \text{N}_d(\boldsymbol\mu_j,\boldsymbol\Sigma_j).
\end{equation*}

$\bullet$ The full conditional distribution of $\theta$ is
\begin{equation*}
\setlength{\jot}{10pt}
\begin{aligned}
\pi(\theta|&\text{rest}) \\
& \propto  \prod_{i=1}^{n}\sqrt{\theta}\text{exp}\Big(-\frac{\theta(Y_i-\boldsymbol Z_{i}^{\top}\boldsymbol\alpha-\boldsymbol E_i^\top \boldsymbol\beta-\kappa_1 \tilde{u_i}) ^2}{\kappa_2^2\tilde{u_i}} \Big)\times \prod_{i=1}^{n}[\theta \text{exp}(-\theta \tilde{u_i})]\times \theta^{a-1}\text{exp}(-b\theta)\\
&\propto \theta^{\frac{3}{2}n+a-1}\text{exp}\Bigg(-\Big(\frac{1}{2}\sum_{i=1}^{n}\frac{(Y_i-\boldsymbol Z_{i}^{\top}\boldsymbol\alpha-\boldsymbol E_i^\top \boldsymbol\beta-\kappa_1 \tilde{u_i}) ^2}{\kappa_2^2\tilde{u_i}} +\sum_{i=1}^{n}\tilde{u_i}+b \Big)\theta \Bigg)
\end{aligned}
\end{equation*}
Therefore, 
\begin{equation*}
\theta|\text{rest} \thicksim \text{Gamma}\Big(\frac{3}{2}n+a, \frac{1}{2}\sum_{i=1}^{n}\frac{(Y_i-Z_{i}^{\top}\alpha-E_i^\top \beta-\kappa_1 \tilde{u_i}) ^2}{\kappa_2^2\tilde{u_i}} +\sum_{i=1}^{n}\tilde{u_i}+b  \Big).
\end{equation*}

$\bullet$ The full conditional distribution of $\eta^2$ is
\begin{equation*}
\setlength{\jot}{10pt}
\begin{aligned}
\pi(\eta^2|\text{rest})
& \propto  \prod_{j=1}^{p}(\frac{\eta^{2}}{2})^{\frac{d+1}{2}}\text{exp}(-\frac{\eta^{2}}{2}g_j )\times (\eta^2)^{c-1} \text{exp}(-m \eta^2)\\
&\propto (\eta^{2})^{\frac{(d+1)p}{2}+c-1}\text{exp}(-(\frac{1}{2}\sum_{j=1}^{p}g_j+m )\eta^2)
\end{aligned}
\end{equation*}
It follows that 
\begin{equation*}
\eta^2|\text{rest}\thicksim\text{Gamma}\Big(\frac{(d+1)p}{2}+c, \frac{1}{2}\sum_{j=1}^{p}g_j+m\Big).
\end{equation*}

$\bullet$ The full conditional distribution of $\boldsymbol\beta$ is
\begin{equation*}
\setlength{\jot}{10pt}
\begin{aligned}
\pi(\boldsymbol\beta|&\text{rest}) \\
& \propto \prod_{i=1}^{n}\text{exp}\Big(-\frac{\theta}{2\kappa_2^2 \tilde{u_i}}(\boldsymbol Y_i-\sum_{j=0}^{p}\boldsymbol \alpha_j ^\top\boldsymbol Z_{ij}-\boldsymbol E_i^\top \boldsymbol\beta-\kappa_1 \tilde{u_i} )^2\Big)\text{exp}(-\frac{1}{2}\boldsymbol\beta^\top \boldsymbol\Sigma_{\boldsymbol\beta}^{-1}\boldsymbol\beta) \\
& \propto \text{exp}\Bigg(-\frac{1}{2}\Big(\boldsymbol\beta^\top(\sum_{i=1}^{n}\frac{\theta \boldsymbol E_i \boldsymbol E_i^\top}{\kappa_2^2 \tilde{u_i}}+\boldsymbol\Sigma_{\boldsymbol\beta}^{-1})\boldsymbol\beta-2\sum_{i=1}^{n}\frac{\theta}{\kappa_2^2 \tilde{u_i}}(\boldsymbol Y_i-\sum_{j=0}^{p}\boldsymbol \alpha_j ^\top\boldsymbol Z_{ij}-\kappa_1\tilde{u_i})\boldsymbol E_i^\top \boldsymbol\beta \Big) \Bigg),
\end{aligned}
\end{equation*}
therefore, we have
\begin{equation*}
\boldsymbol\beta|\text{rest} \thicksim \text{N}_q (\boldsymbol\mu_{\beta^{\star}},\boldsymbol\Sigma_{\beta^{\star}}),
\end{equation*}
with mean
\begin{equation*}
\boldsymbol\mu_{\beta^{\star}}=\boldsymbol\Sigma_{\beta^{\star}}\Big(\sum_{i=1}^{n}\frac{\theta}{\kappa_2^2 \tilde{u_i}}(Y_i-\sum_{j=0}^{p}\boldsymbol \alpha_j ^\top\boldsymbol Z_{ij}-\kappa_1\tilde{u_i})\boldsymbol E_i^\top \Big)^\top
\end{equation*}
and covariance
\begin{equation*}
\boldsymbol\Sigma_{\beta^{\star}}=(\sum_{i=1}^{n}\frac{\theta \boldsymbol E_i \boldsymbol E_i^\top}{\kappa_2^2 \tilde{u_i}}+\boldsymbol\Sigma_{\boldsymbol\beta}^{-1})^{-1}.
\end{equation*}
$\bullet$ The full conditional distribution of $\boldsymbol\alpha_0$ is derived as
\begin{equation*}
\setlength{\jot}{10pt}
\boldsymbol\alpha_0|\text{rest} 
\thicksim \text{N}_d(\boldsymbol\mu_0,\boldsymbol\Sigma_0),
\end{equation*}
where 
\begin{equation*}
\boldsymbol\Sigma_0 = (\sum_{i=1}^{n}\frac{\theta \boldsymbol Z_{i0} \boldsymbol Z_{i0}^\top}{\kappa_2^2 \tilde{u_i}}+\boldsymbol\Sigma_{\boldsymbol\alpha_0}^{-1})^{-1}
\end{equation*}
and 
\begin{equation*}
\boldsymbol\mu_0 = \boldsymbol\Sigma_0\Big(\sum_{i=1}^{n}\frac{\theta}{\kappa_2^2 \tilde{u_i}}(Y_i-\boldsymbol E_i^\top \boldsymbol\beta-\sum_{j=1}^{p}\boldsymbol \alpha_j ^\top\boldsymbol Z_{ij}-\kappa_1\tilde{u_i})\boldsymbol Z_{i0}^\top \Big)^\top.
\end{equation*}

\subsection{Posterior inference for BVCSS}

\subsubsection{Bayesian hierarchical model}

\begin{equation*}
\boldsymbol Y|\boldsymbol\beta,\boldsymbol\alpha,\sigma^2 \thicksim \text{N}_n(\boldsymbol E\boldsymbol\beta+\boldsymbol Z\boldsymbol\alpha,\sigma^2\textbf{I}_n),
\end{equation*}
\begin{equation*}
\boldsymbol\alpha_j|\zeta_j^2,\sigma^2 \text{ } \overset{\mathrm{ind}}{\thicksim} \text{ }   (1-\pi_0)\text{N}_d(0,\sigma^2\zeta_j^2\textbf{I}_d)+\pi_0\delta_0(\boldsymbol\alpha_j), j=1,...,p,
\end{equation*}
\begin{equation*}
\zeta_j^2|\lambda^2 \text{ } \overset{\mathrm{ind}}{\thicksim} \text{ }  \text{Gamma}(\frac{d+1}{2},\frac{\lambda^2}{2}), j=1,...,p,
\end{equation*}
\begin{equation*}
\pi_0 \thicksim \text{Beta}(a,b),
\end{equation*}
\begin{equation*}
\sigma^2 \thicksim \text{Inverse-Gamma}(s,h),
\end{equation*}
\begin{equation*}
\lambda^2 \thicksim  \text{Gamma}(t,\psi),
\end{equation*}
\begin{equation*}
\boldsymbol \beta \thicksim \text{N}_q(0,\boldsymbol \Sigma_{\boldsymbol \beta}),
\end{equation*}
\begin{equation*}
\boldsymbol\alpha_0 \thicksim \text{N}_d(0,\boldsymbol\Sigma_{\boldsymbol\alpha_0}).
\end{equation*}

\subsubsection{Gibbs Sampler}
$\bullet$ The full conditional distribution of $\boldsymbol \alpha_j$, $j=1,...,p$,
\begin{equation*}
\setlength{\jot}{10pt}
\begin{aligned}
\pi(\boldsymbol \alpha_j|\text{rest}) 
& \propto \text{exp}(-\frac{1}{2\sigma^2}||\boldsymbol  Y-\boldsymbol Z_{-j}\boldsymbol \alpha_{-j}-\boldsymbol Z_j\alpha_j-\boldsymbol E\boldsymbol \beta||^2)\\
&\times \Bigg((1-\pi_0)(2\pi \sigma^2\zeta_j^2)^{-\frac{d}{2}}\text{exp}\Big(-\frac{1}{2}\boldsymbol \alpha_j^\top(\sigma^2\zeta_j^2\textbf{I}_d)^{-1}\boldsymbol \alpha_j \Big)\textbf{I}_{(\boldsymbol \alpha_j \neq 0)}+\pi_0 \delta_0 (\boldsymbol \alpha_j)\Bigg) 
\end{aligned}
\end{equation*}

Let $l_j=p(\boldsymbol\alpha_j=0|\text{rest})$, then the conditional posterior distribution of $\boldsymbol\alpha_j(j=1,...,p)$ is a multivariate spike-and-slab distribution given as:
\begin{equation*}
\boldsymbol\alpha_j|\text{rest} \thicksim (1-l_j)\text{N}_d(\boldsymbol\mu_j,\sigma^2\boldsymbol\Sigma_j) + l_j\delta_0(\boldsymbol\alpha_j),
\end{equation*}
where $	\boldsymbol\Sigma_j=(\boldsymbol Z_j^\top\boldsymbol Z_j +\zeta_j^{-2}\textbf{I}_d)^{-1}$, $\boldsymbol \mu_j=\boldsymbol \Sigma_j \boldsymbol Z_j^\top(\boldsymbol Y-\boldsymbol E\boldsymbol \beta-\boldsymbol Z_{-j}\boldsymbol \alpha_{-j})$, and
\begin{equation*}
l_j=\frac{\pi_0}{\pi_0+(1-\pi_0)( \zeta_j^2)^{-\frac{d}{2}}\sqrt{|\boldsymbol \Sigma_j|}\text{exp}\Big(\frac{1}{2}\boldsymbol \mu_j^\top (\sigma^2 \boldsymbol \Sigma_j)^{-1}\boldsymbol \mu_j  \Big)}.
\end{equation*}
Hence, the posterior distribution of $\boldsymbol\alpha_j$ is a mixture of a multivariate normal distribution and a point mass at 0.

$\bullet$ The full conditional distribution of $\sigma^2$
\begin{equation*}
\setlength{\jot}{10pt}
\begin{aligned}
\pi(\sigma^2|&\text{rest}) \\
& \propto (\sigma^2)^{-\frac{n}{2}}\text{exp}\Big(-\frac{1}{2\sigma^2}||\boldsymbol Y-\boldsymbol Z\boldsymbol \alpha-\boldsymbol E\boldsymbol \beta||^2\Big)\times(\frac{1}{\sigma^2})^{s+1}\text{exp}(-\frac{h}{\sigma^2})\\
&\times \prod_{j=1}^{p}\Bigg((1-\pi_0)(2\pi \sigma^2\zeta_j^2)^{-\frac{d}{2}}\text{exp}\Big(-\frac{1}{2}\alpha_j^\top(\sigma^2\zeta_j^2\textbf{I}_d)^{-1}\boldsymbol \alpha_j \Big)\textbf{I}_{(\boldsymbol \alpha_j \neq 0)}+\pi_0 \delta_0 (\boldsymbol \alpha_j)\Bigg) 
\end{aligned}
\end{equation*}
Let 
\begin{equation*}
Q_j=\begin{cases}
\scalebox{1}{0}& { \text{if} \; \boldsymbol \alpha_{j} = 0} \\[6pt]
\scalebox{1}{1}& { \text{if} \; \boldsymbol \alpha_{j} \neq 0}
\end{cases}
\end{equation*}
then the posterior distribution of $\sigma^2$ becomes
\begin{equation*}
\setlength{\jot}{10pt}
\begin{aligned}
\pi(\sigma^2|&\text{rest}) \\
& \propto (\sigma^2)^{-\frac{n}{2}}\text{exp}\Big(-\frac{1}{2\sigma^2}||\boldsymbol Y-\boldsymbol Z\boldsymbol \alpha-\boldsymbol E\boldsymbol \beta||^2\Big)\times(\frac{1}{\sigma^2})^{s+1}\text{exp}(-\frac{h}{\sigma^2})\\
&\times \prod_{j=1}^{p}(1-\pi_0)^{Q_j}(\sigma^2)^{-\frac{d}{2}\sum_{j=1}^{p}Q_j}\prod_{j=1}^{p}\pi_0^{1-Q_j}\text{exp}\Big(-\frac{1}{\sigma^2}\cdot\frac{1}{2}\sum_{j=1}^{p}(\zeta_j^2)^{-1}\boldsymbol \alpha_j^\top\boldsymbol \alpha_j \Big)\\
& \propto (\sigma^2)^{-\frac{n}{2}-\frac{d}{2}\sum_{j=1}^{p}Q_j-s-1}\text{exp}\Big(-\frac{1}{\sigma^2}\Big(\frac{1}{2}||\boldsymbol Y-\boldsymbol Z\boldsymbol \alpha-\boldsymbol E\boldsymbol \beta||^2+\frac{1}{2}\sum_{j=1}^{p}(\zeta_j^2)^{-1}\boldsymbol \alpha_j^\top\boldsymbol \alpha_j +h\Big)\Big).
\end{aligned}
\end{equation*}
Therefore, 
\begin{equation*}
\sigma^2|\text{rest}\thicksim\text{Inverse-Gamma}\Big(\frac{n}{2}+\frac{d}{2}\sum_{j=1}^{p}Q_j+s,\frac{1}{2}||\boldsymbol Y-\boldsymbol Z\boldsymbol \alpha-\boldsymbol E\boldsymbol \beta||^2+\frac{1}{2}\sum_{j=1}^{p}(\zeta_j^2)^{-1}\boldsymbol \alpha_j^\top\boldsymbol \alpha_j +h \Big).
\end{equation*}

$\bullet$ The full conditional distribution of $\zeta_j^2$, $j=1,...,p$,
\begin{equation*}
\setlength{\jot}{10pt}
\begin{aligned}
\pi(\zeta_j^2|&\text{rest})\\
& \propto \Bigg((1-\pi_0)(2\pi \sigma^2\zeta_j^2)^{-\frac{d}{2}}\text{exp}\Big(-\frac{1}{2}\boldsymbol \alpha_j^\top(\sigma^2\zeta_j^2\textbf{I}_d)^{-1}\boldsymbol \alpha_j \Big)\textbf{I}_{(\boldsymbol \alpha_j \neq 0)}+\pi_0 \delta_0 (\boldsymbol \alpha_j)\Bigg)\\
& \times (\zeta_j^2)^{\frac{d-1}{2}}\text{exp}(-\frac{\lambda^2}{2}\zeta_j^2).
\end{aligned}
\end{equation*}

Then we have
\begin{equation*}
(\zeta_j^2)^{-1}|\text{rest} \thicksim \begin{cases}
\scalebox{1}{Inverse-Gamma($\frac{d+1}{2}$,\, $\frac{\lambda^{2}}{2}$)}& { \text{if} \; \boldsymbol \alpha_{j} = 0} \\[6pt]
\scalebox{1}{Inverse-Gaussian($\sqrt{\frac{\sigma^2\lambda^2}{\boldsymbol \alpha_j^\top \boldsymbol \alpha_j}}$,$\lambda^{2}$)}& { \text{if} \; \boldsymbol \alpha_{j} \neq 0}
\end{cases}
\end{equation*}
$\bullet$ The full conditional distribution of $\lambda^2$
\begin{equation*}
\setlength{\jot}{10pt}
\begin{aligned}
\pi(\lambda^2|\text{rest})
& \propto \prod_{j=1}^{p}\Big((\frac{\lambda^2}{2})^{\frac{d+1}{2}}\text{exp}(-\frac{\lambda^2}{2}\zeta_j^2)\Big) \times (\lambda^2)^{t-1}\text{exp}(-\psi \lambda^2)\\
& \propto (\lambda^2)^{\frac{1}{2}(d+1)p+t-1}\text{exp}\Big(-(\frac{1}{2}\sum_{j=1}^{p}\zeta_j^2+\psi )\lambda^2\Big),
\end{aligned}
\end{equation*}
and we have
\begin{equation*}
\lambda^2|\text{rest}\thicksim \text{Gamma}(\frac{1}{2}(d+1)p+t,\frac{1}{2}\sum_{j=1}^{p}\zeta_j^2+\psi).
\end{equation*}

$\bullet$ The full conditional distribution of $\pi_0$
\begin{equation*}
\setlength{\jot}{10pt}
\begin{aligned}
\pi(\pi_0|&\text{rest}) \\
& \propto \prod_{j=1}^{p}\Bigg((1-\pi_0)(2\pi \sigma^2\zeta_j^2)^{-\frac{d}{2}}\text{exp}\Big(-\frac{1}{2}\boldsymbol \alpha_j^\top(\sigma^2\zeta_j^2\textbf{I}_d)^{-1}\boldsymbol \alpha_j \Big)\textbf{I}_{(\boldsymbol \alpha_j \neq 0)}+\pi_0 \delta_0 (\boldsymbol \alpha_j)\Bigg) \\
&\times \pi_0^{a-1}(1-\pi_0)^{b-1}\\
& \propto \pi_0^{a+p-\sum_{j=1}^{p}Q_j-1}(1-\pi_0)^{b+\sum_{j=1}^{p}Q_j-1} ,
\end{aligned}
\end{equation*}
hence
\begin{equation*}
\pi_0|\text{rest}\thicksim\text{Beta}(p+a-\sum_{j=1}^{p}Q_j,b+\sum_{j=1}^{p}Q_j).
\end{equation*}

$\bullet$ The full conditional distribution of $\boldsymbol \beta$
\begin{equation*}
\setlength{\jot}{10pt}
\begin{aligned}
\pi(\boldsymbol \beta|\text{rest}) 
& \propto \text{exp}\Big(-\frac{1}{2\sigma^2}||\boldsymbol Y-\boldsymbol E\boldsymbol \beta-\boldsymbol Z\boldsymbol \alpha||^2\Big)\times \text{exp}(-\frac{1}{2}\boldsymbol \beta^\top \boldsymbol \Sigma_{\boldsymbol \beta}^{-1}\boldsymbol \beta)\\
& \propto \text{exp}\Bigg(-\frac{1}{2}\Big(\boldsymbol \beta^\top(\frac{\boldsymbol E^\top \boldsymbol E}{\sigma^2}+\boldsymbol \Sigma_{\boldsymbol \beta}^{-1})\boldsymbol \beta- \frac{2}{\sigma^2}(\boldsymbol Y-\boldsymbol Z\boldsymbol \alpha)^\top \boldsymbol E\boldsymbol \beta \Big) \Bigg),
\end{aligned}
\end{equation*}
and 
\begin{equation*}
\boldsymbol \beta|\text{rest}\thicksim \text{N}_q(\boldsymbol\mu_{\beta^{\star}},\boldsymbol\Sigma_{\beta^{\star}})
\end{equation*}
where $\boldsymbol\Sigma_{\beta^{\star}}=\Big(\frac{\boldsymbol E^\top \boldsymbol E}{\sigma^2}+\boldsymbol \Sigma_{\boldsymbol \beta}^{-1}\Big)^{-1}$ and $	\boldsymbol\mu_{\beta^{\star}}=\boldsymbol\Sigma_{\beta^{\star}}\Big(\frac{1}{\sigma^2}(\boldsymbol Y-\boldsymbol Z\boldsymbol \alpha)^\top \boldsymbol E\Big)^\top$.

$\bullet$ The full conditional distribution of $\boldsymbol\alpha_0$ is 
\begin{equation*}
\setlength{\jot}{10pt}
\boldsymbol\alpha_0|\text{rest} 
\thicksim \text{N}_d(\boldsymbol\mu_0,\boldsymbol\Sigma_0),
\end{equation*}
with $\boldsymbol\Sigma_0 = \Big(\frac{\boldsymbol Z_{0}^\top \boldsymbol Z_{0}}{\sigma^2}+\boldsymbol\Sigma_{\boldsymbol\alpha_0}^{-1}\Big)^{-1}$ and $\boldsymbol\mu_0 = \boldsymbol\Sigma_0\Big(\frac{1}{\sigma^2 }(Y-\boldsymbol E \boldsymbol\beta-\boldsymbol Z_{-0}\boldsymbol \alpha_{-0})^\top\boldsymbol Z_{0} \Big)^\top$.

\subsection{Posterior inference for BVC}

\subsubsection{Bayesian hierarchical model}

\begin{equation*}
\boldsymbol Y|\boldsymbol \beta,\boldsymbol \alpha,\sigma^2,\zeta_j^2 \thicksim \text{N}_n(\boldsymbol E\boldsymbol \beta+\boldsymbol Z\boldsymbol \alpha,\sigma^2\textbf{I}_n),
\end{equation*}
\begin{equation*}
\boldsymbol \alpha_j|\zeta_j^2,\sigma^2 \thicksim  \text{N}_d(0,\sigma^2\zeta_j^2\textbf{I}_d), j=1,...,p,
\end{equation*}
\begin{equation*}
\zeta_j^2|\lambda^2 \thicksim \text{Gamma}(\frac{d+1}{2},\frac{\lambda^2}{2}), j=1,...,p,
\end{equation*}
\begin{equation*}
\sigma^2 \thicksim \text{Inverse-Gamma}(s,h),
\end{equation*}
\begin{equation*}
\lambda^2 \thicksim \text{Gamma}(t,\psi),
\end{equation*}
\begin{equation*}
\boldsymbol\beta \thicksim \text{N}_q(0,\boldsymbol\Sigma_{\boldsymbol\beta}),
\end{equation*}
\begin{equation*}
\boldsymbol\alpha_0 \thicksim \text{N}_d(0,\boldsymbol\Sigma_{\boldsymbol\alpha_0}).
\end{equation*}

\subsubsection{Gibbs Sampler}
$\bullet$ The full conditional distribution of $\boldsymbol \alpha_j$, $j=1,...,p$
\begin{equation*}
\setlength{\jot}{10pt}
\begin{aligned}
p(\boldsymbol \alpha_j|&\text{rest}) \\
& \propto \text{exp}(-\frac{1}{2\sigma^2}||\boldsymbol Y-\boldsymbol Z\boldsymbol \alpha-\boldsymbol E\boldsymbol \beta||^2)\text{exp}\Big(-\frac{1}{2}\boldsymbol \alpha_j^\top(\sigma^2\zeta_j^2\textbf{I}_d)^{-1}\boldsymbol \alpha_j \Big)\\
& \propto \text{exp}\Bigg(-\frac{1}{2\sigma^2}\Big(\boldsymbol \alpha_j^\top \boldsymbol Z_j^\top \boldsymbol Z_j \boldsymbol \alpha_j-2\boldsymbol \alpha_j \boldsymbol Z_j^\top(\boldsymbol Y-\boldsymbol E\boldsymbol \beta-\boldsymbol Z_{-j}\boldsymbol \alpha_{-j}) \Big) \Bigg)\text{exp}\Big(-\frac{1}{2}\boldsymbol \alpha_j^\top(\sigma^2\zeta_j^2\textbf{I}_d)^{-1}\boldsymbol \alpha_j \Big)\\
& \propto \text{exp}\Bigg(-\frac{1}{2\sigma^2}\Big(\boldsymbol \alpha_j^\top (\boldsymbol Z_j^\top \boldsymbol Z_j +\zeta_j^{-2}\textbf{I}_d)\boldsymbol \alpha_j-2\boldsymbol \alpha_j \boldsymbol Z_j^\top(\boldsymbol Y-\boldsymbol E\boldsymbol \beta-\boldsymbol Z_{-j}\boldsymbol \alpha_{-j}) \Big) \Bigg),
\end{aligned}
\end{equation*}
Denote $
\boldsymbol \Sigma_j=(\boldsymbol Z_j^\top \boldsymbol Z_j +\zeta_j^{-2}\textbf{I}_d)^{-1}$ and $
\boldsymbol \mu_j=\boldsymbol \Sigma_j \boldsymbol Z_j^\top(\boldsymbol Y-\boldsymbol E\boldsymbol \beta-\boldsymbol Z_{-j}\boldsymbol \alpha_{-j})$,
then the posterior distribution of $\boldsymbol \alpha_j$ is
\begin{equation*}
\boldsymbol \alpha_j|\text{rest}\thicksim\text{N}_d(\boldsymbol \mu_j,\sigma^2 \boldsymbol \Sigma_j), \quad j=1,...p.
\end{equation*}

$\bullet$ The full conditional distribution of $\boldsymbol \beta$
\begin{equation*}
\setlength{\jot}{10pt}
\begin{aligned}
p(\boldsymbol \beta|\text{rest}) 
& \propto \text{exp}(-\frac{1}{2\sigma^2}||\boldsymbol Y-\boldsymbol E\boldsymbol \beta-\boldsymbol Z\boldsymbol \alpha||^2)\times \text{exp}(-\frac{1}{2}\boldsymbol \beta^\top \boldsymbol \Sigma_{\boldsymbol \beta}^{-1}\boldsymbol \beta)\\
& \propto \text{exp}\Bigg(-\frac{1}{2}\Big(\boldsymbol \beta^\top(\frac{\boldsymbol E^\top \boldsymbol E}{\sigma^2}+\boldsymbol \Sigma_{\boldsymbol \beta}^{-1})\boldsymbol \beta- \frac{2}{\sigma^2}(\boldsymbol Y-\boldsymbol Z\boldsymbol \alpha)^\top \boldsymbol E\boldsymbol \beta \Big) \Bigg),
\end{aligned}
\end{equation*}
and we have
\begin{equation*}
\boldsymbol \beta|\text{rest}\thicksim \text{N}_q(\boldsymbol\mu_{\beta^{\star}},\boldsymbol\Sigma_{\beta^{\star}})
\end{equation*}
which is a multivariate normal distribution, with mean
\begin{equation*}
\boldsymbol\mu_{\beta^{\star}}=((\frac{\boldsymbol E^\top \boldsymbol E}{\sigma^2}+\boldsymbol \Sigma_{\boldsymbol \beta}^{-1})^{-1}(\frac{1}{\sigma^2}(\boldsymbol Y-\boldsymbol Z\boldsymbol \alpha)^\top \boldsymbol E)^\top
\end{equation*}
and covariance
\begin{equation*}
\boldsymbol\Sigma_{\beta^{\star}}=(\frac{\boldsymbol E^\top \boldsymbol E}{\sigma^2}+\boldsymbol \Sigma_{\boldsymbol \beta}^{-1})^{-1}.
\end{equation*}

$\bullet$ The full conditional distribution of $\zeta_j^2$, $j=1,...,p$
\begin{equation*}
\setlength{\jot}{10pt}
\begin{aligned}
p(\zeta_j^2|\text{rest}) 
& \propto (2\pi \sigma^2\zeta_j^2)^{-\frac{d}{2}}\text{exp}\Big(-\frac{1}{2}\boldsymbol \alpha_j^\top(\sigma^2\zeta_j^2\textbf{I}_d)^{-1}\boldsymbol \alpha_j \Big)\times (\zeta_j^2)^{\frac{d-1}{2}}\text{exp}(-\frac{1}{2}\lambda^2\zeta_j^2)
\\
& \propto (\zeta_j^2)^{-\frac{1}{2}}\text{exp}(-\frac{1}{2}(\frac{\boldsymbol \alpha_j^\top \boldsymbol \alpha_j}{\sigma^2}\frac{1}{\zeta_j^2}+\lambda^2\zeta_j^2)),
\end{aligned}
\end{equation*}
therefore $(\zeta_j^2)^{-1}\thicksim \text{Inverse-Gaussian}(\sqrt{\frac{\sigma^2\lambda^2}{\boldsymbol \alpha_j^\top \boldsymbol \alpha_j}},\lambda^2)$.

$\bullet$ The full conditional distribution of $\lambda^2$
\begin{equation*}
\setlength{\jot}{10pt}
\begin{aligned}
p(\lambda^2|\text{rest}) 
& \propto \prod_{j=1}^{p}\Big((\frac{\lambda^2}{2})^{\frac{d+1}{2}}\text{exp}(-\frac{\lambda^2}{2}\zeta_j^2)\Big) \times (\lambda^2)^{t-1}\text{exp}(-\psi \lambda^2)\\
& \propto (\lambda^2)^{\frac{1}{2}(d+1)p+t-1}\text{exp}(-(\frac{1}{2}\sum_{j=1}^{p}\zeta_j^2+\psi )\lambda^2),
\end{aligned}
\end{equation*}
then,
\begin{equation*}
\lambda^2|\text{rest}\thicksim \text{Gamma}(\frac{1}{2}(d+1)p+t,\frac{1}{2}\sum_{j=1}^{p}\zeta_j^2+\psi).
\end{equation*}

$\bullet$The full conditional distribution of $\sigma^2$
\begin{equation*}
\setlength{\jot}{10pt}
\begin{aligned}
p(\sigma^2|&\text{rest}) \\
& \propto (\sigma^2)^{-\frac{n}{2}}\text{exp}(-\frac{1}{2\sigma^2}||\boldsymbol Y-\boldsymbol Z\boldsymbol \alpha-\boldsymbol E\boldsymbol \beta||^2)\times(\frac{1}{\sigma^2})^{s+1}\text{exp}(-\frac{h}{\sigma^2})\\
&\times \prod_{j=1}^{p}(2\pi \sigma^2\zeta_j^2)^{-\frac{d}{2}}\text{exp}\Big(-\frac{1}{2}\boldsymbol \alpha_j^\top(\sigma^2\zeta_j^2\textbf{I}_d)^{-1}\boldsymbol \alpha_j \Big)\\
& \propto (\sigma^2)^{-\frac{n}{2}-\frac{d(p+1)}{2}-s-1}\text{exp}\Big(-\frac{1}{\sigma^2}(\frac{1}{2}||\boldsymbol Y-\boldsymbol Z\boldsymbol \alpha-\boldsymbol E\boldsymbol \beta||^2+\frac{1}{2}\sum_{j=1}^{p}(\zeta_j^2)^{-1}\boldsymbol \alpha_j^\top\boldsymbol \alpha_j +h)\Big)
\end{aligned}
\end{equation*}
Therefore, the posterior distribution of $\sigma^2$ is
\begin{equation*}
\sigma^2|\text{rest}\thicksim\text{Inverse-Gamma} \Big(\frac{n+dp}{2}+s,\frac{1}{2}||\boldsymbol Y-\boldsymbol Z\boldsymbol \alpha-\boldsymbol E\boldsymbol \beta||^2+\frac{1}{2}\sum_{j=1}^{p}(\zeta_j^2)^{-1}\boldsymbol \alpha_j^\top\boldsymbol \alpha_j +h \Big).
\end{equation*}

$\bullet$ The full conditional distribution of $\boldsymbol\alpha_0$ is derived as
\begin{equation*}
\setlength{\jot}{10pt}
\boldsymbol\alpha_0|\text{rest} 
\thicksim \text{N}_d(\boldsymbol\mu_0,\boldsymbol\Sigma_0),
\end{equation*}
where $	\boldsymbol\Sigma_0 = (\frac{\boldsymbol Z_{0}^\top \boldsymbol Z_{0}}{\sigma^2}+\boldsymbol\Sigma_{\boldsymbol\alpha_0}^{-1})^{-1}$ and 
$	\boldsymbol\mu_0 = \boldsymbol\Sigma_0\Big(\frac{1}{\sigma^2 }(Y-\boldsymbol E \boldsymbol\beta-\boldsymbol Z_{-0}\boldsymbol \alpha_{-0})^\top\boldsymbol Z_{0} \Big)^\top$.

\end{document}